\documentclass[papersize,journal]{IEEEtran}
\usepackage{xcolor}
\usepackage{amsmath}
\usepackage{amsmath,amsfonts}
\usepackage{algorithmic}
\usepackage{algorithm}
\usepackage{array}
\usepackage{subcaption}
\usepackage[caption=false,font=normalsize,labelfont=sf,textfont=sf]{subfig}
\usepackage{textcomp}
\usepackage{stfloats}
\usepackage{url}
\usepackage{verbatim}
\usepackage{graphicx}
\usepackage{cite}
\usepackage{amsthm,amsmath,amssymb,lipsum}
\usepackage{mathrsfs}
\usepackage{makecell}
\usepackage{threeparttable}
\usepackage{amssymb}
\usepackage{multirow}
\usepackage{flushend}
\usepackage{amsmath,amsfonts}
\usepackage{epstopdf}
\usepackage{textcomp}
\usepackage[normalem]{ulem}
\begin{document}
\title{Zak-Transform-Induced Optimal Sequences and Their Applications in OTFS}
\author{Xiuping Peng,\IEEEmembership{ Member, IEEE}, Congying Wu, Zilong Liu, \IEEEmembership{Senior Member, IEEE}, Chunlei Li, \IEEEmembership{Senior Member, IEEE}, Jianye Zhang\IEEEmembership{}, Xiangjun Li, Pingzhi Fan, \IEEEmembership{Fellow, IEEE}


 \thanks{Xiuping Peng, Congying Wu and Jianye Zhang are with the School of Information Science and Engineering, Yanshan University, Qinhuangdao, China, and also with Hebei Key Laboratory of Information Transmission and Signal Processing, Qinhuangdao, China. (e-mail: pengxp@ysu.edu.cn; wu1999109@163.com; 1403224729@qq.com)}

\thanks{Zilong Liu is with the School of Computer Science and Electronic Engineering, University of Essex, UK. (e-mail: zilong.liu@essex.ac.uk)}
  
\thanks{Chunlei Li is with the Department of Informatics,
University of Bergen, 5020 Bergen, Norway. (e-mail: chunlei.li@uib.no)}

\thanks{Pingzhi Fan and Xiangjun Li are with the Information Coding \& Transmission Key Lab of Sichuan Province, CSNMT Int. Coop. Res. Centre (MoST), Southwest Jiaotong University, Chengdu 61175, China. (e-mail: lxj@my.swjtu.edu.cn;p.fan@ieee.org)}}


\markboth{Journal of \LaTeX\ Class Files,~Vol.~14, No.~8, August~2024}%
{Shell \MakeLowercase{\textit{et al.}}: A Sample Article Using IEEEtran.cls for IEEE Journals}


\maketitle

\begin{abstract}
This paper introduces a novel finite Zak transform (FZT)-aided framework for constructing multiple zero-correlation zone (ZCZ) sequence sets with optimal correlation properties. Specifically, each sequence is perfect with zero auto-correlation sidelobes, each ZCZ sequence set meets the Tang-Fan-Matsufuji bound with equality, and the maximum inter-set cross-correlation of multiple sequence sets meets the Sarwate bound with equality. Our study shows that these sequences can be sparsely expressed in the Zak domain through properly selected index and phase matrices. Particularly, it is found that the maximum inter-set cross-correlation beats the Sarwate bound if every index matrix is a circular Florentine array. Several construction methods of multiple ZCZ sequence sets are proposed, demonstrating both the optimality and high flexibility. {Additionally, it is shown that excellent synchronization performance can be achieved by the proposed sequences in orthogonal-time-frequency-space (OTFS) systems.}
\end{abstract}

\begin{IEEEkeywords}Perfect sequences, Zak transform, multiple ZCZ sequence sets,   Sarwate bound, cyclically distinct, inter-set cross-correlation.
\end{IEEEkeywords}

\section{Introduction}
\subsection{Background}
\IEEEPARstart{S}{equences} with good correlation properties are useful for a number of applications (e.g., synchronization, channel estimation, spread-spectrum communication, random access, ranging and positioning) in communication and radar systems. {To deal with  asynchronous wireless channels, perfect sequences with zero auto-corelation sides are preferred. However, perfect binary and quaternary sequences are only known to have lengths of 4 and 2, 4, 8, 16, respectively \cite{Luke2003}. There are polyphase perfect sequences of lengths $N=sm^2$ ($s,m$ positive integers), but as conjectured by Mow in \cite{Mow1996}, their minimum alphabet size is $2sm$ for even $s$ and odd $m$ and is $sm$ otherwise. Furthermore, constrained by the Sarwate bound \cite{Sarwate79}, it is not possible to have two or more perfect sequences with zero cross-correlation functions. }

As a remedy to the aforementioned problem, zero-correlation zone (ZCZ) sequences \cite{Fan99} have received tremendous research attention in the past decades. By definition, ZCZ sequences are characterized by their zero auto- and cross- correlation values for certain time-shifts around the in-phase position. Thanks to this property, ZCZ sequences permit an interference-free window, thus leading to improved multi-user detection or channel estimation performance in, for example, quasi-synchronous code-division multiple access (QS-CDMA) communications \cite{ref2,ref6} or multiple-antenna transmissions \cite{Yang02,Fragouli03,Fan04,Yang08,Zhang12}, respectively.

Formally, let us consider a ZCZ sequence set of length $N$, set size of $T$, the ZCZ width of $Z$. Such a set is featured by their zero periodic (non-trivial) auto- and cross-correlation functions for all the time-shifts in the range of $\left| \tau  \right| < Z$. The Tang-Fan-Matsufuji bound \cite{Tang00} shows that the parameters of a ZCZ sequence set should satisfy $T{Z} \le N$. A ZCZ sequence set is said to be \textit{optimal} if it meets this bound with equality. To support multi-cell QS-CDMA or multi-user MIMO communications, there is a strong need to design multiple ZCZ sequence sets having low inter-set cross-correlation with respect to the Sarwate bound \cite{Sarwate79} or the generalised Sarwate bound for the binary case \cite{Peng80}. 

\subsection{Related Works}
As a class of orthogonal design, a number of ZCZ constructions from various aspects have been developed. Typically, one can design ZCZ sequences from perfect sequences (e.g., generalized Chirp-like sequences) as illustrated in \cite{Torii04,Hayashi09,Popovic10}. Hu and Gong \cite{Hu10} presented a general construction of sequence families with zero or low correlation zones using interleaving techniques and Hadamard matrices. Besides, the research works of \cite{Deng00} and \cite{Appuswamy06} showed that complementary sequences \cite{Golay61,Tseng72,Liu13,Liu14} are an useful building component of ZCZ sequences. The algebraic connection between mutually orthogonal complementary sets and ZCZ sequences through generalized Reed-Muller codes was revealed in \cite{Liu14-ITW}. 

Designing multiple ZCZ sets with low inter-set cross-correlation \cite{Tang10} is a challenging task. \cite{Zhou17} pointed out that multiple ZCZ sequence sets with optimal inter-set cross-correlation can also be obtained by extending the method in \cite{Popovic10}. The resultant multiple ZCZ sequence sets have the following properties: 1) each sequence is perfect with zero auto-correlation sidelobes; 2) each ZCZ sequence set meets the Tang-Fan-Matsufuji bound with equality; and 3) the maximum inter-set cross-correlation of multiple sequence sets meets the Sarwate bound with equality. However, some of the ZCZ sequences obtained in \cite{Popovic10} may be cyclically equivalent, which is not desirable in practical applications \cite{Golomb2005}. To solve this problem, improved multiple ZCZ sets were obtained with the aid of perfect nonlinear functions \cite{Zhou17} and generalized bent functions \cite{Zhang20}. Recently, circular Florentine arrays were employed in \cite{Zhang22} for more ZCZ sequence sets compared to that in \cite{Zhou17},\cite{Zhang20}. The same combinatorial tool was used in \cite{Song21} for sequences with perfect auto-correlation and optimal cross-correlation.

\subsection{Motivations and Contributions}
Against the above state-of-the-art, this paper seeks a novel research angle for new optimal multiple ZCZ sequence sets. We advocate the use of an emerging tool, called finite Zak transform (FZT), which has found wide applications in mathematics, quantum mechanics, and signal analysis \cite{ref18,ref19,ref20}. A key advantage of FZT is that the sparse representation of sequences in Zak space enables efficient signal processing in radar, sonar, and communications \cite{ref21,ref22}, leading to reduced computational complexity as well as storage space at the receiver. Building upon FZT and its inverse, Brodzik derived sequences with perfect auto-correlation \cite{ref24} and all-zero cross-correlation \cite{ref25}. Recently, FZT was utilized in \cite{Peng24} for multiple spectrally-constrained sequence sets with optimal ZCZ and all-zero inter-set cross-correlation properties. Yet, the full potential of FZT for sequence design is largely unexplored. 

{From the application perspective, owing to the equivalence between the Zak domain and the delay-Doppler (DD) domain, FZT has inspired orthogonal-time-frequency-space (OTFS) modulation, which is a promising multicarrier waveform for future high-mobility communications \cite{6GV2X}. In the first version of OTFS, the basic idea is to send the data symbols in the DD domain (i.e., Zak domain), convert it to time-frequency (TF) domain through inverse sympletic finite Fourier transform (ISFFT), and then to time domain via Heisenberg transform \cite{Hadani2017,Viterbo2018}. Recently, it has been found that one can directly generate the relevant time-domain signal by applying the inverse FZT (IFZT) to the DD domain data \cite{Saif2022, Saif2023,Lampel2022}. Therefore, it is intriguing to transmit the proposed Zak-transform-induced sequences in the DD domain as preamble sequences and investigate their performances for random access \cite{Viterbo2020}, synchronization \cite{Cho2021, Farhang2022,Chung2024}, channel estimation \cite{Yogesh2024}, sensing \cite{Zegrar2024}, etc.} 

The contributions of this work are multifold:
\begin{itemize}
    \item We first introduce a novel framework whereby optimal multiple ZCZ sequence sets can be uniquely obtained by IFZT. To this end, we introduce a number of index matrices and phase matrices by advocating the sequence sparsity in the Zak domain. 
    \item We derive the admissible conditions of these index matrices and phase matrices and show that the maximum inter-set cross-correlation beats the Sarwate bound when every index matrix is a circular Florentine array. 
     \item {We employ the proposed sequences as preamble sequences in the DD domain and study their synchronization performance in OTFS. Our numerical simulation results demonstrate that 1) their superiority over random sequences in OTFS synchronization and 2) their excellent ambiguity function, highlighting their potential for sensing. }
\end{itemize}

\subsection{Organization of This Work}
The rest of the paper is organized as follows. Section II gives brief introductions to perfect sequences, FZT, and cyclic Florentine arrays. For optimal multiple ZCZ sequence sets, we introduce the main framework and derive the conditions for index and phase matrices in the Zak domain in Section III. In Section IV, several constructions are proposed based on IFZT and the cyclic Florentine arrays. The derived sequences are then applied to an OTFS system and evaluated for their synchronization performance. Section V concludes this paper.

\section{Preliminaries}
\subsection{Perfect Sequences}
\textit{Definition 1:} Let ${\textbf{s}_0}=\left( {{s_0\left( 0 \right)},...,{s_0\left( N-1 \right)}} \right)$ and ${\textbf{s}_1}=\left( {{s_1\left( 0 \right)},...,{s_1\left( N-1 \right)}} \right)$ be two sequences of  period $N$, then the periodic cross-correlation function (PCCF) between ${\textbf{s}_0}$ and ${\textbf{s}_1}$ is defined as 
\begin{equation}
{{\theta _{\textbf{s}_0,\textbf{s}_1}}\left( \tau  \right) = \sum\limits_{n = 0}^{N - 1} {s_0\left( n+\tau\right)_N{s_1^*}{{\left( n  \right)}}},} 
\end{equation}
where $0 \le \tau  < N$, ${\left(  \cdot  \right)_N}$ indicates the integer modulus of $N$  and
$s_1^*\left(n\right)$ is the complex conjugate of the complex number $s_1\left(n\right)$. When $\textbf{s}_0 = \textbf{s}_1$, ${\theta_{\textbf{s}_0}}$ is called the periodic auto-correlation function (PACF) of $\textbf{s}_0$. A sequence $\textbf{s}_0$ is said to be perfect if ${\theta _{\textbf{s}_0}}\left( \tau  \right)=0 $ for all $0 < \tau  < N$.

{\textit{Definition 2:} For two sequences ${\textbf{s}_0}$ and ${\textbf{s}_1}$ with period $N$, if there exist some $0 \le \tau  < N$ and a constant complex number $c$ with $|c|=1$ such that $s_1(t)=cs_0(t+\tau)$ for all $0 \le t  < N$ (i.e., $|R_{s_0,s_1}(\tau)|=N$), then the sequences ${\textbf{s}_0}$ and ${\textbf{s}_1}$ are said to be cyclically equivalent. Otherwise, they are said to be cyclically distinct.}

\textit{Definition 3:} Let $S = \left\{ {{\textbf{s}_u}} \right\}_{u= 0}^{T - 1}$ be a set of $T$ sequences of period $N$, where ${\textbf{s}_u}=\left( {{s_u\left( 0 \right)},...,{s_u\left( N-1 \right)}} \right)$ denotes the $u$-th constituent sequence of $S$. The maximum out-of-phase periodic auto-correlation magnitude ${\theta _a}$ and the maximum periodic cross-correlation magnitude ${\theta _c}$ of the sequence set $S$ are respectively defined by
\begin{equation*}
{\theta _a} = \max \left\{ {\left| {{\theta _{\textbf{s}_u}}\left( \tau  \right)} \right|:0 \le u< T,0 < \tau  < N} \right\},
\end{equation*}
and
\begin{equation*}
{\theta _c} = \max \left\{ {\left| {{\theta _{{{\rm{\textbf{s}}}_u}{\rm{,}}{{\rm{\textbf{s}}}_v}}}\left( \tau  \right)} \right|:0 \le u \ne v < T,0 \le \tau  < N} \right\}.
\end{equation*}

The following lemma is the well-known Sarwate bound on ${\theta _a}$ and ${\theta _c}$.

\textit{Lemma 1 \cite{Sarwate79}:} For any sequence set $S$ with $T$ sequences of period $N$, we have
\begin{equation}
\frac{{\theta _c^2}}{N} + \frac{{N - 1}}{{N\left( {T - 1} \right)}}\frac{{\theta _a^2}}{N} \ge 1.
\end{equation}

Lemma 1 demonstrates that it is impossible to obtain a sequence set with both $\theta _a$ and $\theta _c$ being zero. This implies that cross-correlation and nontrivial auto-correlation cannot be zero for all $\tau$. Fortunately, this problem can be addressed by placing $\tau$ in some regions around the origin, which facilitates the application and development of sequences with zero correlation zones \cite{Fan99}.

\textit{Definition 4:} The set $S$ is called an $\left( {N,T,{Z}} \right)$-ZCZ sequence set if 
\begin{equation*}
\begin{aligned}
{\theta _{\textbf{s}_u}}\left( \tau  \right) = 0,~&for ~ 0 \le u < T~and~0 < \tau  < {Z},\\
{\theta _{{\textbf{s}_u},{\textbf{s}_v}}}\left( \tau  \right) = 0, ~&for~0 \le u \ne v < T~ and~0 \le \tau  < {Z},
\end{aligned}
\end{equation*}
where ${Z}$ denotes the length of the ZCZ.

The following bound implies that there is a tradeoff among the parameters of any ZCZ sequence set.

\textit{Lemma 2 (Tang-Fan-Matsufuji bound \cite{Tang00}):} Let $S$ be a set of $T$ sequences of period $N$ with ZCZ length $Z$, then
\begin{equation*}
T{Z} \le N.
\end{equation*}

A ZCZ sequence set meeting the Tang-Fan-Matsufuji bound is said to be optimal.

\textit{Definition 5:} Let ${{\cal S}}$ be a family of $M$ sequence sets, each consisting of $T$ sequences of period $N$, i.e., ${{\cal S}} = \left\{ {{S^0},{S^1},...{S^{M - 1}}} \right\}$. A sequence set $S^m (0 \le m< M)$ is expressed as:
\begin{equation*}
{S^m} = \left\{ {{\textbf{s}}_u^m:{\textbf{s}}_u^m \buildrel \Delta \over = \left\{ {s_u^m\left( n \right)} \right\}_{n = 0}^{{N} - 1},0 \le u < T} \right\},
\end{equation*}
The inter-set cross-correlation of ${{\cal S}}$ is defined as
 \[{\theta_c}\left( {{\cal S}} \right)= \max \left\{ {\left| {{\theta _{\textbf{s}_u^{{m_1}},\textbf{s}_v^{{m_2}}}}\left( \tau  \right)} \right|:0 \le \tau  < N }\right\}\]
where $0 \le {m_1} \ne {m_2} <M$ and $0 \le u, v <T$.

\subsection{The Zak Transform}
\textit{Definition 6 \cite{ref24}:} Let ${\textbf{s}}$ be a sequence of  period $N$. Suppose that $N = LT$, where $L$ and $T$ are positive integers. The FZT of ${\textbf{s}}$ is given by
\begin{equation}
\label{deqn_ex1}
{X}(j,t) = \sum\limits_{l = 0}^{L - 1} {s(t + lT)w_L^{-lj}},0 \le j < L, 0 \le t < T,
\end{equation}
where ${w_L} = {e^{\frac{{2\pi \sqrt { - 1} }}{L}}} = {e^{\frac{{2\pi i}}{L}}}$.

When a sequence of period $N=LT$ is represented as an $L\times T$ matrix $x$, its FZT domain ${X}$ can be rewritten as ${X}=W \cdot x$, where $W=w_L^{-lj}$ is the discrete fourier transform (DFT) matrix of order $L$. It is clear that the FZT reduces to the classic DFT when $T=1$.

Similarly to the DFT, the FZT is a one-to-one mapping. A signal  ${\textbf{s}}$ can be recovered by its ${X}(j,t)$ as
\begin{equation}
\label{deqn_ex2}
\textbf{s}(t + lT) = {L^{ - 1}}\sum\limits_{j = 0}^{L - 1} {{X}(j,t)w_L^{lj},}
\end{equation}
where $0 \le t < T$ and $0 \le l < L$.

Take ${X}$, ${Y}$ and ${Z}$ as the FZTs of $\textbf{s}_0$, $\textbf{s}_1$ and ${\theta _{\textbf{s}_0,\textbf{s}_1}}$, respectively. The Zak space correlation formula is given by
\begin{equation}
{Z}\left( {j,t} \right) =\sum\limits_{k = 0}^{T- 1} {{X}(j,k+t)Y^*(j,k).} 
\end{equation} 

Consequently, for a shift $\tau = \tau_1+\tau_2T$ where $0\leq \tau_1 <T$ and $0\leq \tau_2 < L$, we have
\begin{equation}
\begin{aligned}
&{\theta _{{\bf{s}}_0,{\bf{s}}_1}}\left( { {\tau_1} +  {\tau_2}T} \right)\\
=& {L^{ - 1}}\sum\limits_{j = 0}^{L - 1} \sum\limits_{t = 0}^{T - 1} {X\left( {j,t+\tau_1} \right)Y{^*}\left( {j,t} \right){w_L^{ {\tau_2}j}} } 
\end{aligned}
\end{equation}



\subsection{Circular Florentine Arrays}
The circular Florentine array has been studied since 1989 \cite{ref27,ref28,ref29}. The definition and some lemmas about circular Florentine arrays are introduced in the following.

{\textit{Definition 7 \cite{ref30}:} An $M \times T$ circular Florentine array is an array of $T$ distinct symbols in $M$ circular rows  such that 
\begin{enumerate}
\item{each row is a permutation of the $T$ symbols and }
\item{for any pair of distinct symbols $\left( {s,t} \right)$ and for each $1 \le a \le T- 1$, there is at most one row in which $t$ occurs $a$ steps to the right of $s$.}
\end{enumerate}
}
\textit{Example 1:} An example of a $4\times15$ circular Florentine array is shown in (7).
 \begin{figure*}[ht] 
 	\centering
 	\begin{equation}	
\left[ {
\setlength{\arraycolsep}{7pt}
\begin{array}{*{20}{c}}
0&1&2&3&4&5&6&7&8&9&{10}&{11}&{12}&{13}&{14}\\
0&7&1&8&2&{12}&3&{11}&9&4&{13}&5&{14}&6&{10}\\
0&4&{11}&7&{10}&1&{13}&9&5&8&3&6&2&{14}&{12}\\
0&{13}&7&2&{11}&6&{14}&{10}&3&5&{12}&9&1&4&8
\end{array}} \right].
\end{equation}
\end{figure*}

\textit{Lemma 3 }: For each positive integer $T \ge 2$, let ${F_c}\left( T \right)$ denote the largest integer such that circular Florentine array of order ${F_c}\left( T \right) \times T$ exists, then we have
\begin{enumerate}
\item{$s - 1 \le {F_c}\left( T \right) \le T - 1$, where $s$ is the smallest prime factor of $T$\cite{ref30};}
\item{${F_c}\left( T \right) =1$ when $T$ is even \cite{ref31};}
\item{${F_c}\left( T \right) =T-1$ when $T$ is prime \cite{ref30}; and}
\item{${F_c}\left( T \right) \le T-3$ when $T \equiv 15\bmod 18$ \cite{ref30}.}
\end{enumerate}

The following lemma of the circular Florentine array guarantees the cross-correlation properties of the sequence sets, which can be used in the proof of Lemma 8. 

\textit{Lemma 4 \cite{ref30}:} Let ${\mathbb{Z}_T} = \left\{ {0,1,...,T - 1} \right\}$, $F$ be an ${F_c}\left( T \right) \times T$ circular Florentine array on  ${\mathbb{Z}_T}$. Then each row is an arrangement on ${\mathbb{Z}_T}$, denoted as ${\pi _i}$, where $0 \le i < {F_c}\left( T \right)$. For $0 \le {i_1} \ne {i_2} \le{F_c}\left( T \right) $ and $z \in{\mathbb{Z}_T} $, there is exactly one solution for ${\pi _{{i_1}}}\left( t \right) = {\pi _{{i_2}}}\left( {t + z} \right)$ on ${\mathbb{Z}_T}$.

\section{Proposed Zak-Transform-Induced Multiple ZCZ Sequence Sets}
{In this section, we present a novel Zak-transform-induced framework for constructing multiple ZCZ sequence sets. Our proposed framework advocates the sparse representations of these sequences in the Zak domain. Our key idea is that sequences within a set exhibit identical non-zero support in the Zak domain, whilst distinct sets possess different non-zero supports. Following this idea, we first identify the Zak-domain non-zero positions of each set using an \textit{index matrix}. Subsequently, we assign the corresponding Zak-domain phase values to these non-zero positions, represented by a \textit{phase matrix}. In short, the proposed framework is comprised of three steps: 1) determining the Zak-domain non-zero positions using the index matrix; 2) assigning the Zak-domain phase values using the phase matrix; and 3) generating time-domain sequences of a set via the IFZT. To generate multiple good sequence sets, appropriate index matrices and phase matrices are needed.}


{\textit{Main Framework: } Let $R$, $M$, $T$ and $L$ be positive integers. 
$M$ multiple sequence sets ${{\cal S}} = \left\{ {S^m} \right\}_{m = 0}^{M - 1}$, each comprising $T$ sequences of period $N$, where $N=RT^2=LT$, are constructed by following the steps below.
\begin{enumerate}
\item{Let $A$ be an index matrix, which is a $M \times T$ matrix over $\mathbb{Z}_T$. 
$A^m\left( t \right)$ denotes the value of the $t$-th element in the $m$-th row of the index matrix $A$. The row vector $A^m$ for $0\leq m <M$ corresponds to the sequence set $S^m$. 
The non-zero support of the sequence set $S^m$ in the Zak domain is given by $\left| {X^{m}}(j,t) \right|$ as follows: 
\begin{equation}
\begin{aligned}
\left| {{X^{m}}\left( {j,t} \right)} \right| = \begin{cases}T\sqrt R , &j =A^m\left( t \right) + {r}T,\\ 0, & { \text{otherwise} }\end{cases}
\end{aligned}
\end{equation}
where $0 \le j < L,0 \le t< T$ and $0\le r<R$. }
\item{Let $P^m (0\le m<M)$ be a phase matrix, which is a $T \times L$ matrix. ${P_u^m}\left( t \right)$ denotes the value of the $t$-th element in the $u$-th row of the phase matrix $P^m$. The row vector $P_u^m$ corresponds to the sequence ${\textbf{s}}_u^m$ in $S^m$. 
The sequence ${\textbf{s}}_u^m$ in the Zak domain is given by ${X_{u}^{m}}\left( {j,t} \right)$ as follows:}
\begin{equation}
\begin{aligned}
{{X_{u}^{m}}\left( {j,t} \right)} = \begin{cases}T\sqrt R {P^m_u}\left( {t + {r}T} \right), &j = A^m\left( t \right)  +{r}T\\ 0, &{\text{otherwise.} }\end{cases}
\end{aligned}
\end{equation}
\item{For each $0 \le m < M$, according to the IFZT,  the sequence $s_u^m(n)$ in $S^m$ is obtained by
\begin{equation}
\begin{aligned}
&s_u^m\left( n \right)= s_u^m\left( {t + lT} \right) \\
= &{L^{ - 1}}T\sqrt R \sum\limits_{{r} = 0}^{R-1} {P_u^m\left( {t + {r}T} \right)w_L^{ l\left( {A^m\left( t \right) + {r}T} \right)}},
\end{aligned}
\end{equation}
where $0 \le u<T$ and $0 \le l < L$.}
\end{enumerate}
{To demonstrate the construction of multiple sequence sets using the index matrix $A$ and the phase matrices $P^m (0\le m<M)$, an example of the proposed Main Framework is provided for the case where $T=5$, $M=R=2$ and $L=RT=10$.}
}
\textit{Example 2:} $A$ is a $2\times 5$ index matrix, which is expressed as:
\[A= \left[ {\begin{array}{*{20}{c}}
0&1&2&4&3\\
0&3&4&2&1\\
\end{array}} \right].\]

 \begin{figure*}[ht] 
 	\centering
\begin{subequations}
\begin{align}
P^0 &= \left[ {\begin{array}{*{20}{c}}
{w_{10}^0}&{w_{10}^0}&{w_{10}^0}&{w_{10}^0}&{w_{10}^0}&{w_{10}^0}&{w_{10}^0}&{w_{10}^0}&{w_{10}^0}&{w_{10}^0}\\
{w_{10}^0}&{w_{10}^1}&{w_{10}^2}&{w_{10}^3}&{w_{10}^4}&{w_{10}^5}&{w_{10}^6}&{w_{10}^7}&{w_{10}^8}&{w_{10}^9}\\
{w_{10}^0}&{w_{10}^2}&{w_{10}^4}&{w_{10}^6}&{w_{10}^8}&{w_{10}^0}&{w_{10}^2}&{w_{10}^4}&{w_{10}^6}&{w_{10}^8}\\
{w_{10}^0}&{w_{10}^3}&{w_{10}^6}&{w_{10}^9}&{w_{10}^2}&{w_{10}^5}&{w_{10}^8}&{w_{10}^1}&{w_{10}^4}&{w_{10}^7}\\
{w_{10}^0}&{w_{10}^4}&{w_{10}^8}&{w_{10}^2}&{w_{10}^6}&{w_{10}^0}&{w_{10}^4}&{w_{10}^8}&{w_{10}^2}&{w_{10}^6}
\end{array}} \right], \tag{\theequation a}\\
P^1 &= \left[ {\begin{array}{*{20}{c}}
{w_{10}^0}&{w_{10}^5}&{w_{10}^0}&{w_{10}^5}&{w_{10}^0}&{w_{10}^5}&{w_{10}^0}&{w_{10}^5}&{w_{10}^0}&{w_{10}^5}\\
{w_{10}^0}&{w_{10}^6}&{w_{10}^2}&{w_{10}^8}&{w_{10}^4}&{w_{10}^0}&{w_{10}^6}&{w_{10}^2}&{w_{10}^8}&{w_{10}^4}\\
{w_{10}^0}&{w_{10}^7}&{w_{10}^4}&{w_{10}^1}&{w_{10}^8}&{w_{10}^5}&{w_{10}^2}&{w_{10}^9}&{w_{10}^6}&{w_{10}^3}\\
{w_{10}^0}&{w_{10}^8}&{w_{10}^6}&{w_{10}^4}&{w_{10}^2}&{w_{10}^0}&{w_{10}^8}&{w_{10}^6}&{w_{10}^4}&{w_{10}^2}\\
{w_{10}^0}&{w_{10}^9}&{w_{10}^8}&{w_{10}^7}&{w_{10}^6}&{w_{10}^5}&{w_{10}^4}&{w_{10}^3}&{w_{10}^2}&{w_{10}^1}
\end{array}} \right]. \tag{\theequation b}
\end{align}
\end{subequations}
 \end{figure*}
Let $P^0$ and $P^1$ be the phase matrices, as shown in (11a) and (11b), respectively. Through (8) and (9), we can get ${{X_{1}^{0}}\left( {j,t} \right)}$ and ${{X_{3}^{1}}\left( {j,t} \right)}$, i.e.,
\[X_{1}^{0}= 5\sqrt2\left[ {
\begin{array}{*{20}{c}}
{w_{10}^0}&0&0&0&0\\
0&{w_{10}^1}&0&0&0\\
0&0&{w_{10}^2}&0&0\\
0&0&0&0&{w_{10}^4}\\
0&0&0&{w_{10}^3}&0\\
{w_{10}^5}&0&0&0&0\\
0&{w_{10}^6}&0&0&0\\
0&0&{w_{10}^7}&0&0\\
0&0&0&0&{w_{10}^9}\\
0&0&0&{w_{10}^8}&0
\end{array}} \right],\]
\[X_{3}^{1}=  5\sqrt2\left[ {
\begin{array}{*{20}{c}}
{w_{10}^0}&0&0&0&0\\
0&0&0&0&{w_{10}^2}\\
0&0&0&{w_{10}^4}&0\\
0&{w_{10}^8}&0&0&0\\
0&0&{w_{10}^6}&0&0\\
{w_{10}^0}&0&0&0&0\\
0&0&0&0&{w_{10}^2}\\
0&0&0&{w_{10}^4}&0\\
0&{w_{10}^8}&0&0&0\\
0&0&{w_{10}^6}&0&0
\end{array}} \right].\]

{In this paper, the proposed Main Framework will be employed to construct multiple sequence sets with the following desired properties:}
\begin{enumerate}
\item{Each sequence is a perfect unimodular sequence;}
\item{Each $S^m$ is an optimal ZCZ sequence set with respect to the Tang-Fan-Matsufuji bound;}
\item{The family of sequence set ${\cal S}$ has low inter-set cross-correlation, namely, the maximal inter-set cross-correlation of multiple sequence sets achieves the well-known Sarwate bound;}
\item{All sequences in each $S^m$ are cyclically distinct.}
\end{enumerate}

{The above analysis has revealed that the key to constructing multiple sequence sets with the aforementioned properties is the design of appropriate index matrix $A$ and its corresponding phase matrices $P^m(0\le m<M-1)$. To proceed, we first introduce the necessary conditions that such matrices must satisfy. These conditions play a pivotal role in the construction of multiple sequence sets with desirable properties, which will be detailed in Section IV.}

\textit{Lemma 5:} Let $\textbf{s}^m_u$ be a sequence of  period $N$, where $N=LT$ and $ L=RT$. The sequence $\textbf{s}^m_u$ is unimodular if the phase vector $P_u^m$ of $\textbf{s}^m_u$ in Zak domain satisfies the following condition: 
\begin{equation}
\left| {\sum\limits_{{r} = 0}^{R - 1} {{P^m_u}\left( {t + {r}T} \right)w_L^{  lrT}} } \right| = \sqrt R 
\end{equation}
for all $0 \le t<T$ and $0 \le l <L$.  

\textit{Proof:} To ensure that $\left| {s_u^m\left( n \right)} \right|=1$ holds, from (10), the time-domain expression of $\textbf{s}^m_u$ is
\begin{equation*}
\begin{aligned}
&s_u^m\left( n \right)= s_u^m\left( {t + lT} \right) \\
= &{L^{ - 1}}T\sqrt R \sum\limits_{{r} = 0}^{R-1} {{P^m_u}\left( {t + {r}T} \right)w_L^{  l\left( {A^m\left( t \right) + {r}T} \right)}} \\
 = &{L^{ - 1}}T\sqrt R w_L^{  lA^m\left( t \right)}\sum\limits_{{r} = 0}^{R-1} {{P^m_u}\left( {t + {r}T} \right)w_L^{  l{r}T}} .
\end{aligned}
\end{equation*}
Then $\left| {s_u^m\left( n \right)} \right|=\frac{1}{{\sqrt R }}\left| {\sum\limits_{{r} = 0}^{R-1} {{P^m_u}\left( {t + {r}T} \right)w_L^{  l{r}T}}} \right|$. Since (12) holds for all $0 \le t<T$ and $0 \le l <L$, it follows that each sequence $s_u^m$ is unimodular.
\hfill
$\hfill\blacksquare$ 

\textit{Lemma 6:} Let $\textbf{s}^m_u$ be a sequence of period $N$, where $N=LT$ and $ L=RT$. The sequence $\textbf{s}^m_u$ is perfect if its phase vector $P^m_u$ and index vector $A^m$ in the Zak domain satisfy the following conditions:
\begin{enumerate}
\item{The index vector $A^m$ is a permutation of ${\mathbb{Z}_T}$.}
\item{The phase vector $P^m_u$ is unimodular. }
\end{enumerate}

\textit{Proof:} To ensure that ${\theta _{{\bf{s}}_u^m}}\left( 0 \right) =N$ and ${\theta _{{\bf{s}}_u^m}}\left( \tau  \right) =0$ for all $0 < \tau <N$,  according to (6), the auto-correlation of $\textbf{s}^m_u$ is
\begin{equation}
\begin{aligned}
&{\theta _{{\bf{s}}_u^m}}\left( \tau  \right)={\theta _{{\bf{s}}_u^m}}\left( {{\tau_1} + {\tau_2}T} \right)\\
=&L^{-1}\sum\limits_{j \in {\mathbb{Z}_L}} {w_L^{   {\tau_2}j}\sum\limits_{t \in {\mathbb{Z}_T}} {X_{u}^{m}\left( {j,t+{\tau_1}} \right)X{_{u}^{{m}*}\left( {j,t} \right)} } }
\end{aligned}
\end{equation}
where $0 \le {\tau_1}<T$ and $0 \le {\tau_2} <L$.

When ${\tau_1} \ne0$, since $A^m$ is a permutation of ${\mathbb{Z}_T}$, we get 
\[\sum\limits_{t \in {\mathbb{Z}_T}}{X_{u}^{m}\left( {j,t+{\tau_1}} \right)X{{_{u}^{m}}^*}\left( {j,t} \right)}  = 0.\]
Therefore ${\theta _{{\bf{s}}_u^m}}\left( \tau  \right) = 0$.

When ${\tau_1}=0, {\tau_2}\ne 0$, (13) becomes
\begin{equation*}
\begin{aligned}
&{\theta _{{\bf{s}}_u^m}}\left( { {\tau_2}T} \right) \\
=& L^{-1}\sum\limits_{j \in {\mathbb{Z}_L}} {w_L^{   {\tau_2}j}} \sum\limits_{t \in {\mathbb{Z}_T}} {X_{u}^{m}\left( {j,t} \right)X{{_{u}^{m}}^*}\left( {j,t} \right)} \\
 =& T\sum\limits_{{r} = 0}^{R - 1} {\sum\limits_{t = 0}^{T - 1} {w_L^{   {\tau_2}\left( {{r}T + A^m\left( t\right)} \right)}{P^m_u}\left( {t + {r}T} \right)P_u^{m*}\left( {t + {r}T} \right)} }. 
\end{aligned}
\end{equation*}

Since $A^m$ is a permutation of $\mathbb{Z}_T$, $ A^m\left( t \right) +{r}T$ for $0 \le r<R$ is a permutation of $\mathbb{Z}_L$. Then we have $\sum\limits_{{r} = 0}^{R - 1}\sum\limits_{t = 0}^{T - 1} {w_L^{   {\tau_2}\left( {{r}T + A^m\left( t\right)} \right)}}=0$. Therefore, we can obtain ${\theta _{{\bf{s}}_u^m}}\left( \tau  \right) =0$ for ${\tau_1}=0$ and $0 <{\tau_2}<L$.

When ${\tau_1}=0,  {\tau_2}=0$, (13) becomes
\begin{equation*}
\begin{aligned}
&{\theta _{{\bf{s}}_u^m}}\left( 0\right) \\
 = &T\sum\limits_{{r} = 0}^{R - 1} {\sum\limits_{t = 0}^{T - 1} {w_L^{  0\left( {{r}T + A^m\left( t\right)} \right)}{P^m_u}\left( {t + {r}T} \right)P_u^{m*}\left( {t + {r}T} \right)} } \\
 = &T\sum\limits_{{r} = 0}^{R - 1} {\sum\limits_{t = 0}^{T - 1} {{{\left| {{P^m_u}\left( {t + {r}T} \right)} \right|}^2}} }.
\end{aligned}
\end{equation*}

Since the phase vector $P^m_u$ is unimodular, we can get ${\theta _{{\bf{s}}_u^m}}\left( 0\right) =RT^2=N$. 
\hfill
$\hfill\blacksquare$ 

\textit{Lemma 7:} Let $S^m$ be a set of $T$ sequences of period $N$, where $N=LT$ and $L=RT$. Suppose the phase vector $P^m_u$ and the index vector $A^m$ of the sequence set $S^m$ in the Zak domain satisfy the following conditions:
\begin{enumerate}
\item{The index vector $A^m$ is a permutation of ${\mathbb{Z}_T}$},
\item{$\left| {\sum\limits_{{r} = 0}^{R - 1} {\sum\limits_{t = 0}^{T - 1} {w_L^{  {\tau_2}\left( {A^m\left( t \right) + {r}T} \right)} \cal P} } } \right|=0$, where $0\le  {\tau_2} <R$},
\item{
$0<\left| {\sum\limits_{{r} = 0}^{R - 1} {\sum\limits_{t = 0}^{T - 1} { w_T^{  A^m\left( t \right) } \cal P} }}  \right|<RT,$}
\end{enumerate}
where $\mathcal{P} = P_u^m\left( {t + {r}T} \right)P_v^{m*}\left( {t + {r}T} \right)$.
Then the sequence set $S^m$ satisfies the Tang-Fan-Matsufuji bound and each sequence is cyclically distinct.

\textit{Proof:} Let ${\textbf{s}_u^m}$ and ${\textbf{s}_v^m}$ be any two sequences in ${S^m}$, $0 \le u\ne v < T$ and $0 \le m <M$. In order to ensure that the length of the ZCZ meets Tang-Fan-Matsufuji bound, we need to show $\left| {{\theta _{{\bf{s}}_u^m,{\bf{s}}_v^m}}\left( \tau  \right)} \right|=0$ for $0 \le \tau<RT$. 

Meanwhile, to further ensure that all the sequences are pairwise cyclically distinct, it is sufficient to guarantee $0 < \left| {{\theta _{{\bf{s}}_u^m,{\bf{s}}_v^m}}\left( RT \right)} \right|<N$. {There are two reasons for this constraint:
(1) If $ \left| {{\theta _{{\bf{s}}_u^m,{\bf{s}}_v^m}}\left( RT \right)} \right|=N$, the resulting sequence set must contain equivalent sequences. 
(2) If $ \left| {{\theta _{{\bf{s}}_u^m,{\bf{s}}_v^m}}\left( RT \right)} \right|=0$, then the ZCZ width is $Z=RT+1$, which violates the Tang-Fan-Matsufuji bound.}

According to (6), the cross-correlation of ${\textbf{s}_u^m}$ and ${\textbf{s}_v^m}$ is
\begin{equation}
\begin{aligned}
&{\theta _{{\bf{s}}_u^m,{\bf{s}}_v^m}}\left( \tau  \right) ={\theta _{{\bf{s}}_u^m,{\bf{s}}_v^m}}\left( { {\tau_1} +  {\tau_2}T} \right)\\
=& L^{-1}\sum\limits_{j \in {\mathbb{Z}_L}} {w_L^{   {\tau_2}j}\sum\limits_{t \in {\mathbb{Z}_T}} {X_{u}^{m}\left( {j,t+ {\tau_1}} \right)X{{_{v}^{m}}^*}\left( {j,t} \right)} }, 
\end{aligned}
\end{equation}
where $0 \le  {\tau_1}<T$ and $0 \le {\tau_2} <L$.

When ${\tau_1}\ne0$, since $A^m$ is a permutation of ${\mathbb{Z}_T}$, we can get 
\[\sum\limits_{t \in {\mathbb{Z}_T}}{X_{u}^{m}\left( {j,t+ {\tau_1}} \right)X{{_{v}^{m}}^*}\left( {j,t} \right)}  = 0.\]
Therefore, $\left| {{\theta _{{\bf{s}}_u^m,{\bf{s}}_v^m}}\left( \tau  \right)} \right| = 0$.

When ${\tau_1}=0$, we have the following two cases.

Case 1: When $0 \le  {\tau_2} <R$, (14) becomes
\begin{equation*}
\begin{aligned}
&{\theta _{{\bf{s}}_u^m,{\bf{s}}_v^m}}\left( { {\tau_2}T} \right) \\
=& L^{-1}\sum\limits_{j \in {\mathbb{Z}_L}} {w_L^{   {\tau_2}j}\sum\limits_{t \in {\mathbb{Z}_T}} {X_{u}^{m}\left( {j,t} \right)X{{_{v}^{m}}^*}\left( {j,t} \right)} } \\
 =& T\sum\limits_{{r} = 0}^{R - 1} {\sum\limits_{t = 0}^{T - 1} {w_L^{   {\tau_2}\left( {A^m\left( t\right) + {r}T} \right)}\mathcal{P}} },
\end{aligned}
\end{equation*}
where  ${\mathcal{P}= P_u^m\left( {t + {r}T} \right)P_v^{m*}\left( {t + {r}T} \right)}$.

Given that
\begin{equation}
\left| {\sum\limits_{{r} = 0}^{R - 1} {\sum\limits_{t = 0}^{T - 1} {w_L^{  {\tau_2}\left( {A^m\left( t \right) + {r}T} \right)}\cal P} } } \right|=0,
\end{equation}
 it follows that $\left| {\theta _{{\bf{s}}_u^m,{\bf{s}}_v^m}}\left( {{\tau_2}T} \right) \right| =0$ for all $0 \le {\tau_2} <R$.

Case 2: When ${\tau_2}=R$, (14) becomes
\begin{equation*}
\begin{aligned}
&{\theta _{{\bf{s}}_u^m,{\bf{s}}_v^m}}\left( L \right)\\
 =&L^{-1}\sum\limits_{j \in {\mathbb{Z}_L}} {w_L^{  Rj}\sum\limits_{t \in {\mathbb{Z}_T}} {X_{u}^{m}\left( {j,t} \right)X{{_{v}^{m}}^*}\left( {j,t} \right)} } \\
 = &T\sum\limits_{{r} = 0}^{R - 1} {\sum\limits_{t = 0}^{T - 1} {w_T^{  \left( {A^m\left( t \right) + {r}T} \right)}\mathcal{P}} }\\
= &T\sum\limits_{{r} = 0}^{R - 1} {\sum\limits_{t = 0}^{T - 1} {w_T^{  {A^m\left( t \right)} }\mathcal{P}} },
\end{aligned}
\end{equation*}
where  $\mathcal{P} = P_u^m\left( {t + {r}T} \right)P_v^{m*}\left( {t + {r}T} \right)$.

Given that
\begin{equation}
\begin{aligned}
0<\left| {\sum\limits_{{r} = 0}^{R - 1} {\sum\limits_{t = 0}^{T - 1} { w_T^{  A^m\left( t \right) }\cal {P} } }}  \right|<RT,
\end{aligned}
\end{equation}
 it follows that $0<\left| {{\theta _{{\bf{s}}_u^m,{\bf{s}}_v^m}}\left( {{\tau_2}T} \right)} \right|<N$ for $ {\tau_2}=R$.

Our observations from the two cases demonstrate that the sequence set $S^m$ adheres to the Tang-Fan-Matsufuji bound, while also exhibiting the property of cyclic distinctness for each individual sequence.
\hfill
$\hfill\blacksquare$ 

\textit{Lemma 8:} Consider a family ${\cal S}$ containing $M$ sequence sets, denoted by $S^{m} \left(0 \le {m} < M\right) $, where each set comprises $T$ sequences of period $N$. Here, $N=LT$ and $L=RT$. Let $P^{m_1}$ and $P^{m_2}$ represent the phase matrices, and $A^{m_1}$ and $A^{m_2}$ denote the index vectors associated with sets $S^{m_1}$ and $S^{m_2}$ respectively, where $0 \le {m_1} \ne {m_2} <M$. Then the maximum inter-set cross-correlation of ${{\cal S}}$ attains the Sarwate bound if the following two conditions are met:
\begin{enumerate}
\item{The index matrix $A$ is a circular Florentine array}.
\item{$\left| {\sum\limits_{{r} = 0}^{R - 1} {P_u^{{m_1}}\left( {t +{\tau _1}+{r}T} \right){P_v^{{m_2}*}\left( {t+ {r}T} \right)w_R^{ {r}{\tau _2}}} } }\right| = \sqrt R,$ }
\end{enumerate}
where $0 \le t,\tau _1 < T$, $0 \le u \ne v<T$ and $0 \le \tau _2<L$.

\textit{Proof:} Let ${{\bf{s}}_u^{{m_1}}}$ and ${{\bf{s}}_v^{{m_2}}}$ be any two sequences in $S^{m_1}$ and $S^{m_2}$, respectively. The inter-set cross-correlation between ${{\bf{s}}_u^{{m_1}}}$ and ${{\bf{s}}_v^{{m_2}}}$ is given by
\begin{equation*}
\label{e:barwq}
\begin{aligned}
&{\theta _{{\bf{s}}_u^{{m_1}},{\bf{s}}_v^{{m_2}}}}\left( \tau  \right) = \sum\limits_{n = 0}^{N - 1} {s_u^{{m_1}}\left( {n+ \tau } \right)_Ns{{_v^{{m_2}}}^*}\left( n  \right)} \\
=&\frac{1}{R}\sum\limits_{t = 0}^{T - 1} {\sum\limits_{l = 0}^{L- 1}} \left( {\sum\limits_{{r} = 0}^{R - 1} {P_v^{{m_2}*}\left( {t + {r}T} \right)w_L^{-l\left( {A^{{m_2}}\left( {t } \right) + {r}T} \right)}}} \right. \\
& \left. \cdot \sum\limits_{{r} = 0}^{R - 1} {P_u^{{m_1}}\left( {t + {\tau _1} + {r}T} \right)w_L^{(l+{\tau _2})\left( {A^{{m_1}}\left( t+{\tau _1} \right) + {r}T} \right)}} \right)\\
 =& \frac{1}{R}\sum\limits_{t = 0}^{T - 1} {w_L^{  \tau _2A^{{m_1}}\left( {t + {\tau _1}} \right)}\sum\limits_{l = 0}^{L - 1} {w_L^{l\left(   A^{{m_1}}\left( {t + {\tau _1}}\right)-{A^{{m_2}}\left( t\right) } \right)}} } \\
&{\sum\limits_{{r} = 0}^{R - 1} {P_u^{{m_1}}\left( {t +{\tau _1}+{r}T} \right){P_v^{{m_2}*}\left( {t+ {r}T} \right)w_R^{ {r}{\tau _2}}} } },
\end{aligned}
\end{equation*}
where $n = t + lT$, $\tau  = {\tau _1} + {\tau _2}T$, $0 \le t,{\tau _1} < T$ and $0 \le l,{\tau _2} < L$.

Since
\begin{equation}
\left| {\sum\limits_{{r} = 0}^{R - 1} {P_u^{{m_1}}\left( {t +{\tau _1}+ {r}T} \right){P_v^{{m_2}*}\left( {t + {r}T} \right)w_R^{ {r}{\tau _2}}} }} \right|= \sqrt R,
\end{equation}
the absolute value of the above equation of ${\theta _{{\bf{s}}_u^{{m_1}},{\bf{s}}_v^{{m_2}}}}\left( \tau  \right) $ is 
\begin{equation}
\begin{aligned}
&\left|{\theta _{{\bf{s}}_u^{{m_1}},{\bf{s}}_v^{{m_2}}}}\left( \tau  \right)  \right|\\
=&\frac{\sqrt R}{R}\left|\sum\limits_{t = 0}^{T - 1} {w_L^{  \tau _2A^{{m_1}}\left( {t + {\tau _1}} \right)}\sum\limits_{l = 0}^{L - 1} {w_L^{l\left(   A^{{m_1}}\left( {t + {\tau _1}}\right)-{A^{{m_2}}\left( t\right) } \right)}} } \right| .
\end{aligned}
\end{equation}

Lemma 4 implies that for a circular Florentine array, $A$, there exists a unique integer $t$, denoted $t'$, such that the following equality holds for all possible values of $\tau_1$ within the range $0 \le {\tau_1} <T$:
\[{A^{{m_2}}\left({ t'} \right) = A^{{m_1}}\left( t'+ {\tau _1}  \right)}.\]
 Thus, we have
\[\sum\limits_{l = 0}^{L - 1} {w_L^{l\left( A^{{m_1}}\left( {t'+ {\tau _1} } \right) - {A^{{m_2}}\left({ t' }\right)} \right)}} =\sum\limits_{l = 0}^{L - 1} {1}=L.\]
For distinct values of $t$, i.e. $t \ne t'$, there is no assurance that ${A^{{m_2}}\left( {t } \right)=A^{{m_1}}\left(t+ {\tau _1} \right)}$. In such cases, where ${A^{{m_2}}\left( {t } \right)-A^{{m_1}}\left( t+ {\tau _1} \right)}=b \ne 0$, we obtain
\[\sum\limits_{l = 0}^{L - 1}{w_L^{bl}}=0.\]
Hence, (18) becomes as
\begin{equation*}
\begin{aligned}
&\left|{\theta _{{\bf{s}}_u^{{m_1}},{\bf{s}}_v^{{m_2}}}}\left( \tau  \right)  \right|\\
=&\frac{\sqrt R}{R}{{\left| {w_L^{  {\tau _2}{A^{{m_1}}}\left( {t' + {\tau _1}} \right)}} \right|}L+\frac{\sqrt R}{R} \left| { \sum_{ \substack{t = 0\\t \ne t'}}^{T - 1} {w_L^{  {\tau _2}{A^{{m_1}\left( {t + {\tau _1}} \right)}}} \cdot 0 } }\right|} \\
 =& T\sqrt R.
\end{aligned}
\end{equation*}
Therefore, in light of the preceding discussion and Lemma 1, we can definitively conclude that the maximum inter-set cross-correlation of ${{\cal S}}$ achieves the Sarwate bound.

\section{Sets Of Perfect Sequences With Optimal Correlation}
\subsection{The Proposed Constructions}
Leveraging Lemma 8, we identify that the index matrix possesses the structure of a circular Florentine array. However, current methods for obtaining these arrays primarily rely on computational search techniques. To address this limitation and expand the pool of index matrices available for our constructions, we propose an extension method for circular Florentine arrays. This method offers a more flexible permutation set, facilitating the subsequent construction of multiple optimal ZCZ sequence sets with optimal interset correlation.

\textit{Construction \uppercase\expandafter{\romannumeral1}:} Let $F$ be an ${F_c}\left( T \right) \times T$ circular Florentine array. We denote the $t$-th element in the $m$-th row of $F$ by ${F_m}\left( t \right)$, where ${F_m}\left( t \right)={\left(  t+b \right)_T}$, $0 \le m < {F_c}\left( T \right)$ and $0 \le t,b<T$. We achieve this construction in two steps:
\begin{enumerate}
\item{Choose the first row of $F$, denoted as $F_0$. By rearranging the last $T-2$ elements of ${F_0}$, we obtain a new permutation, denoted as ${F_0^q}$, where $0< q<\left(T-2 \right)!$. 
}
\item{For the $m'$-th row of ${F^q}$ (where $1 \le m' < {F_c}\left( T \right)$), denoted as ${F_{m'}^q}$, the elements can be obtained by
\begin{equation*}
F_{m'}^q\left( t \right) = F_0^q\left( {{F_{m'}}\left( t \right)} \right).
\end{equation*}}
\end{enumerate}
This construction process yields $\left( {T - 2} \right)! - 1$ distinct types of circular Florentine arrays.

\textit{Proof:} We leverage Lemma 4 to establish that for ${m_1} \ne {m_2}$ and $z \in{\mathbb{Z}_T}$, there exists a unique solution for the equation $F_{{m_1}}\left( t \right) = F_{{m_2}}\left( {t + z} \right)$ within ${\mathbb{Z}_T}$. Furthermore, the initial step guarantees that $F_ 0^q$ constitutes a permutation in ${\mathbb{Z}_T}$, and the mapping between the function output $F_0^q\left( t \right) $ and the independent variable $t$ is one-to-one. Consequently, we can assert that there exists a unique solution for the equation $F_0^q\left( {F_{{m_1}}\left( t \right)} \right) = F_0^q\left( {F_{{m_2}}\left( {t + z} \right)} \right)$ within ${\mathbb{Z}_T}$. This implies that $F^q$ itself qualifies as a circular Florentine array.
\hfill
$\hfill\blacksquare$ 

\textit{Remark 1:} Fixing the first two elements in $F_0$ is to ensure that there is no equivalence for all in the extended cases.

\textit{Example 3:} Let $T=5$, the selected circular Florentine array ${F}$, and its $5$ extended circular Florentine arrays $F^q \left( 0<q \le 5 \right)$ obtained by Construction \uppercase\expandafter{\romannumeral1} are listed as follows:
\[\begin{array}{l}
{F}= \left[ {\begin{array}{*{20}{c}}
0&1&2&3&4\\
0&2&4&1&3\\
0&3&1&4&2\\
0&4&3&2&1
\end{array}} \right],{F^1} = \left[ {\begin{array}{*{20}{c}}
0&1&2&4&3\\
0&2&3&1&4\\
0&4&1&3&2\\
0&3&4&2&1
\end{array}} \right],\\
{F^2}= \left[ {\begin{array}{*{20}{c}}
0&1&3&4&2\\
0&3&2&1&4\\
0&4&1&2&3\\
0&2&4&3&1
\end{array}} \right],{F^3} = \left[ {\begin{array}{*{20}{c}}
0&1&3&2&4\\
0&3&4&1&2\\
0&2&1&4&3\\
0&4&2&3&1
\end{array}} \right],\\
{F^4}= \left[ {\begin{array}{*{20}{c}}
0&1&4&3&2\\
0&4&2&1&3\\
0&3&1&2&4\\
0&2&3&4&1
\end{array}} \right],{F^5} = \left[ {\begin{array}{*{20}{c}}
0&1&4&2&3\\
0&4&3&1&2\\
0&2&1&3&4\\
0&3&2&4&1
\end{array}} \right].
\end{array}\]

{Building upon the Main Framework and the insights from Lemmas 5-8 presented in Section \uppercase\expandafter{\romannumeral3}, this section introduces three novel constructions for multiple ZCZ sequence sets with optimal correlation properties for  $T>3$. These constructions are categorized according to the value of $R$: $R=1$, odd $R$, and even $R$. For each case, the desired properties are achieved by carefully designing the index matrix $A$ and the corresponding phase matrices $P^m$.}

\textit{Theorem 1:} Let $R=1$, $M=F_c(T)$ and $N=T^2$. Select the circular Florentine array $F^q \left(0< q<\left(T-2 \right) ! \right)$ as the index matrix $A$, where $F^q$ is obtained by Construction \uppercase\expandafter{\romannumeral1}. The phase matrices $P^m$ for the distinct sequence sets $S^m\left( 0 \le m<M \right)$ are assumed to be identical and collectively denoted as $P$. The ${P_u}\left( t \right)$ of $P$ is defined as:
\begin{equation}
{P_u}\left( t \right)=w_T^{ut},
\end{equation}
where $0\le u,t<T$. According to the Main Framework, ${s_u^m\left( n \right)}$ in $S^m$ is obtained as
\begin{equation}
{s_u^m\left( n \right)}={s}_u^m\left( {t + lT} \right) = {P_u}\left( t \right)w_T^{ lA^m\left( t \right)},
\end{equation}
where $0 \le l < T$. Then the sequence set $S^m\left( 0\le m<M \right)$ has the following properties:
\begin{enumerate}
\item{Each sequence in $S^m$  is unimodular and  perfect.}
\item{Each $S^m$ is an optimal $\left( {{T^2},T,T} \right)$-ZCZ sequence set.}
\item{${\theta _{{\bf{s}}_u^{{m_1}},{\bf{s}}_v^{{m_2}}}}\left( \tau  \right) = T$ for all $0 \le \tau  < T^2,0 \le {m_1} \ne {m_2} < M$ and $0 \le u \ne v <T$.} 
\end{enumerate}

\textit{Proof:}  The detailed proof of Theorem 1 is provided in the Appendix.

Incorporating the circular Florentine array $F$ described in Construction \uppercase\expandafter{\romannumeral1} into the construction framework of Theorem 1 inevitably leads to the issue of sequence equivalence within sequence sets. To circumvent this problem, we propose the following construction method by refining the generation process of the phase matrix.

\textit{Corollary 1:} 
Select the circular Florentine array $F$ as the index matrix $A$. The ${P_u}\left( t \right)$ of $P$ is defined as:
\begin{equation}
\begin{aligned}
{P_u}\left( t \right)= \begin{cases}{w_T^{ut}},& \text{for}~{t = 0,1,\ldots,T - 3,}\\{w_T^{u\left( {T - 1} \right)}}, &\text{for} ~{ t=T-2,}\\{w_T^{u\left( {T -2} \right)}}, &\text{for}~{ t=T-1.}\end{cases}
\end{aligned}
\end{equation}
 The sequence set $S^m\left( 0\le m<M \right)$ obtained by (20) and (21)  has exactly the same properties as Theorem 1.

\textit{Proof:} The proof of Corollary 1 is similar to that for Theorem 1, hence it is omitted.

\textit{Example 4:} Let $N = RT^2$, where $R=1$ and $T=4$. $M={F_c}\left( 4 \right)=1$.  Then we choose the index matrix $A$ as \[{A} = \left[ {\begin{array}{*{20}{c}}
0&1&3&2
\end{array}} \right].\]
From (21), the phase matrix is obtained as
\[P= \left[ {\begin{array}{*{20}{c}}
w_4^0&w_4^0&w_4^0&w_4^0\\
w_4^0&w_4^1&w_4^2&w_4^3\\
w_4^0&w_4^2&w_4^0&w_4^2\\
w_4^0&w_4^3&w_4^2&w_4^1
\end{array}} \right].\]
Then the sequence set $S^0$ consisting of $4$ sequences of period $16$ can be obtained, and the element of $\textbf{s}^0_u $ is expressed as:
\begin{equation*}
{s^0_u\left( n \right)}={s}^0_u\left( {t + 4l} \right) = {{P^0_u}\left( t \right)}w_4^{ lA^0\left( t \right)},
\end{equation*} 
where $0 \le u,t,l < 4$. Four sequences in $S^0$ are given below:
\begin{equation*}
\textbf{s}^0_0 = \left( {1,1,1,1,1,w_4^1,w_4^3,w_4^2,1,w_4^2,w_4^2,1,1,w_4^3,w_4^1,w_4^2} \right) ,
 \end{equation*}
\begin{equation*}
\textbf{s}^0_1  = \left( {1,w_4^1,w_4^2,w_4^3,1,w_4^2,w_4^1,w_4^1,1,w_4^3,1,w_4^3,1,1,w_4^3,w_4^1} \right),
 \end{equation*}
\begin{equation*}
\textbf{s}^0_2 = \left( {1,w_4^2,1,w_4^2,1,w_4^3,w_4^3,1,1,1,w_4^2,w_4^2,1,w_4^1,w_4^1,1} \right),
 \end{equation*}
\begin{equation*}
\textbf{s}^0_3  = \left( {1,w_4^3,w_4^2,w_4^1,1,1,w_4^1,w_4^3,1,w_4^1,1,w_4^1,1,w_4^2,w_4^3,w_4^3} \right).
 \end{equation*}

The PACF of ${\textbf{s}}^0_1$ and PCCF between ${\textbf{s}}^0_2$ and ${\textbf{s}}^0_3$  are shown in Fig. 1. It is known that the sequence set $S^0$ has optimal correlations.
\begin{figure}[!t]
\centering
    \begin{subfigure}{0.48\linewidth}
    \includegraphics[width=\textwidth]{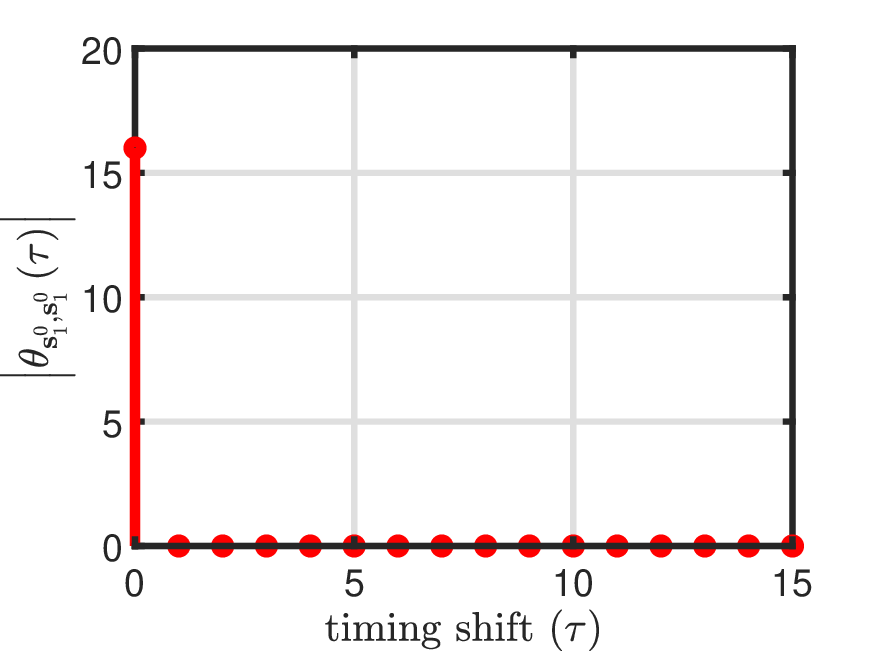}
        \caption{}
        \label{fig_first_case}
    \end{subfigure}
    \begin{subfigure}{0.48\linewidth}
        \includegraphics[width=\textwidth]{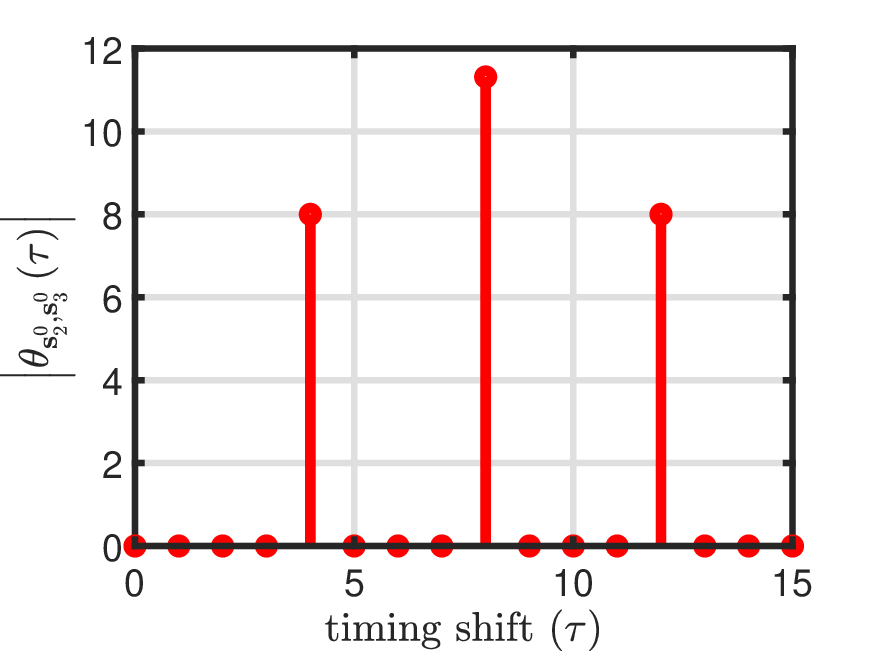}
        \caption{}
        \label{fig_second_case}
    \end{subfigure}    
    \caption{(a) The PACF of ${\textbf{s}}_1^0$ obtained from Example 4; (b) The PCCF of ${\textbf{s}}_2^0$ and ${\textbf{s}}_3^0$.}
    \label{fig_sim}
\end{figure}

\textit{Theorem 2:} Let $R$ be odd, $L=RT$, $N=RT^2$ and $M = \min \left\{ {{R^*} - 1,{F_c}\left( T \right)} \right\}$, where $R^*$ is the smallest prime divisor of $R$. Select $M$ rows randomly from the circular Florentine array $F^q\left(0< q<\left(T-2 \right) ! \right)$ to form the index matrix $A$, where $F^q$ is obtained by Construction \uppercase\expandafter{\romannumeral1}. The ${P^m_u}\left( l\right)$ within the phase matrix $P^m$, associated with the sequence set $S^m$, is defined as:
\[{P_u^m}\left( {l} \right)={P_u^m}\left( {t + {r}T} \right) = w_R^{a^{rm}}w_T^{ut},\]
where $a^{rm}={\left( {m + 1} \right)\frac{{{r}\left( {1 + {r}} \right)}}{2}}$, $0 \le {l} < L$, $0 \le {r} < R$, $0 \le t,u < T$ and $0 \le m < M$. 

According to the Main Framework, ${s_u^m\left( n \right)}$ in $S^m$ is obtained as described in (10).
Then  the sequence set $S^m$, $0 \le m <M$, has the following properties:
\begin{enumerate}
\item{Each sequence in $S^m$ is unimodular and perfect.}
\item{Each $S^m$ is an optimal $\left( {{RT^2},T,RT} \right)$-ZCZ sequence set.}
\item{${\theta _{{\bf{s}}_u^{{m_1}},{\bf{s}}_v^{{m_2}}}}\left( \tau  \right) = \sqrt RT$ for all $0 \le \tau  < RT^2,0 \le {m_1} \ne {m_2} < M$ and $0 \le u \ne v <T$.} 
\end{enumerate}

\textit{Proof:}  The detailed proof of Theorem 2 is provided in the Appendix.


\textit{Corollary 2:} Let $F$ be the circular Florentine array described in Construction \uppercase\expandafter{\romannumeral1}. Select $M$ rows randomly from the circular Florentine array $F$ to form the index matrix $A$. Define a $T\times L$ matrix ${Q^{m}}$, the $l$-th element of the $u$-th row of the $m$-th matrix ${Q^{m}}$ is expressed as 
\[{Q_u^m}\left( {l} \right)={Q_u^m}\left( {t + {r}T} \right) = w_R^{a^{rm}}w_T^{ut},\]
where $a^{rm}={\left( {m + 1} \right)\frac{{{r}\left( {1 + {r}} \right)}}{2}}$, $0 \le {r} < R$, $0 \le t,u < T$ and $0 \le m < M$. The phase matrix $P^m$ is obtained by swapping the $\left( {rT - 1}\right)$-th and $\left( {rT - 2}\right)$-th column of the matrix ${Q^{m}}$
 for $0 < r \le R$. The sequence sets obtained by (10) have exactly the same properties as in Theorem 2.


\textit{Example 5:} Let $N = RT^2$, where $R=3$ and $T=5$. $M = \min \left\{ {3 - 1,{F_c}\left( 5 \right)} \right\}=2$. From Construction \uppercase\expandafter{\romannumeral1}, the constructed circular Florentine array $F^1$ is shown as
 \[{F^1} = \left[ {\begin{array}{*{20}{c}}
0&1&2&4&3\\
0&2&3&1&4\\
0&3&4&2&1\\
0&4&1&3&2
\end{array}} \right].\]
Randomly select the first two rows of array $F^1$ to form the matrix $A$, as shown below:
\[{A} = \left[ {\begin{array}{*{20}{c}}
0&1&2&4&3\\
0&2&3&1&4
\end{array}} \right].\]
The phase matrices $P^0$ and $P^1$ for the sequence sets $S^0$ and $S^1$ in the Zak domain  are shown in (22a) and (22b), respectively.
 \begin{figure*}[ht] 
 	\centering
\begin{subequations}
\begin{align}
P^0 &= \left[ {\begin{array}{*{20}{c}}
{w_{15}^0}&{w_{15}^0}&{w_{15}^0}&{w_{15}^0}&{w_{15}^0}&{w_{15}^5}&{w_{15}^{5}}&{w_{15}^{5}}&{w_{15}^{5}}&{w_{15}^{5}}&{w_{15}^0}&{w_{15}^0}&{w_{15}^0}&{w_{15}^0}&{w_{15}^0}\\
{w_{15}^0}&{w_{15}^3}&{w_{15}^6}&{w_{15}^9}&{w_{15}^{12}}&{w_{15}^{5}}&{w_{15}^8}&{w_{15}^{11}}&{w_{15}^{14}}&{w_{15}^2}&{w_{15}^0}&{w_{15}^3}&{w_{15}^6}&{w_{15}^9}&{w_{15}^{12}}\\
{w_{15}^0}&{w_{15}^6}&{w_{15}^{12}}&{w_{15}^3}&{w_{15}^9}&{w_{15}^5}&{w_{15}^{11}}&{w_{15}^2}&{w_{15}^8}&{w_{15}^{14}}&{w_{15}^0}&{w_{15}^6}&{w_{15}^{12}}&{w_{15}^3}&{w_{15}^9}\\
{w_{15}^0}&{w_{15}^9}&{w_{15}^3}&{w_{15}^{12}}&{w_{15}^6}&{w_{15}^5}&{w_{15}^{14}}&{w_{15}^8}&{w_{15}^2}&{w_{15}^{11}}&{w_{15}^0}&{w_{15}^9}&{w_{15}^3}&{w_{15}^{12}}&{w_{15}^6}\\
{w_{15}^0}&{w_{15}^{12}}&{w_{15}^9}&{w_{15}^6}&{w_{15}^3}&{w_{15}^5}&{w_{15}^2}&{w_{15}^{14}}&{w_{15}^{11}}&{w_{15}^8}&{w_{15}^0}&{w_{15}^{12}}&{w_{15}^9}&{w_{15}^6}&{w_{15}^3}
\end{array}} \right], \tag{\theequation a}\\
P^1 &= \left[ {\begin{array}{*{20}{c}}
{w_{15}^0}&{w_{15}^0}&{w_{15}^0}&{w_{15}^0}&{w_{15}^0}&{w_{15}^{10}}&{w_{15}^{10}}&{w_{15}^{10}}&{w_{15}^{10}}&{w_{15}^{10}}&{w_{15}^0}&{w_{15}^0}&{w_{15}^0}&{w_{15}^0}&{w_{15}^0}\\
{w_{15}^0}&{w_{15}^3}&{w_{15}^6}&{w_{15}^9}&{w_{15}^{12}}&{w_{15}^{10}}&{w_{15}^{13}}&{w_{15}^1}&{w_{15}^4}&{w_{15}^7}&{w_{15}^0}&{w_{15}^3}&{w_{15}^6}&{w_{15}^9}&{w_{15}^{12}}\\
{w_{15}^0}&{w_{15}^6}&{w_{15}^{12}}&{w_{15}^3}&{w_{15}^9}&{w_{15}^{10}}&{w_{15}^1}&{w_{15}^7}&{w_{15}^{13}}&{w_{15}^4}&{w_{15}^0}&{w_{15}^6}&{w_{15}^{12}}&{w_{15}^3}&{w_{15}^9}\\
{w_{15}^0}&{w_{15}^9}&{w_{15}^3}&{w_{15}^{12}}&{w_{15}^6}&{w_{15}^{10}}&{w_{15}^4}&{w_{15}^{13}}&{w_{15}^7}&{w_{15}^1}&{w_{15}^0}&{w_{15}^9}&{w_{15}^3}&{w_{15}^{12}}&{w_{15}^6}\\
{w_{15}^0}&{w_{15}^{12}}&{w_{15}^9}&{w_{15}^6}&{w_{15}^3}&{w_{15}^{10}}&{w_{15}^7}&{w_{15}^4}&{w_{15}^1}&{w_{15}^{13}}&{w_{15}^0}&{w_{15}^{12}}&{w_{15}^9}&{w_{15}^6}&{w_{15}^3}
\end{array}} \right]. \tag{\theequation b}
\end{align}
\end{subequations}

 \end{figure*}
From (10), two sequence sets of size $5$ and period $75$ are obtained as
\begin{equation*}
\begin{aligned}
&s_u^m\left( n \right)= s_u^m\left( {t + 5l} \right)\\
=& \frac{{\sqrt 3 }}{3} \sum\limits_{{r} = 0}^2 {P_u^m\left( {t + 5{r}} \right)w_{15}^{  l\left( {A^m\left( t \right) + 5{r}} \right)}},
\end{aligned}
\end{equation*} 
where $0 \le u,t < 5, 0 \le l<15, 0\le m<2$. The PACF of ${\textbf{s}}^0_1$, the PCCF between ${\textbf{s}}_0^0$ and ${\textbf{s}}_4^0$ and the PCCF between ${\textbf{s}}_0^0$ and ${\textbf{s}}_2^1$ are shown in Fig. 2. 

    
     

\begin{figure}[!h]
    \centering
    \begin{subfigure}{\linewidth}
        \includegraphics[width=0.85\textwidth]{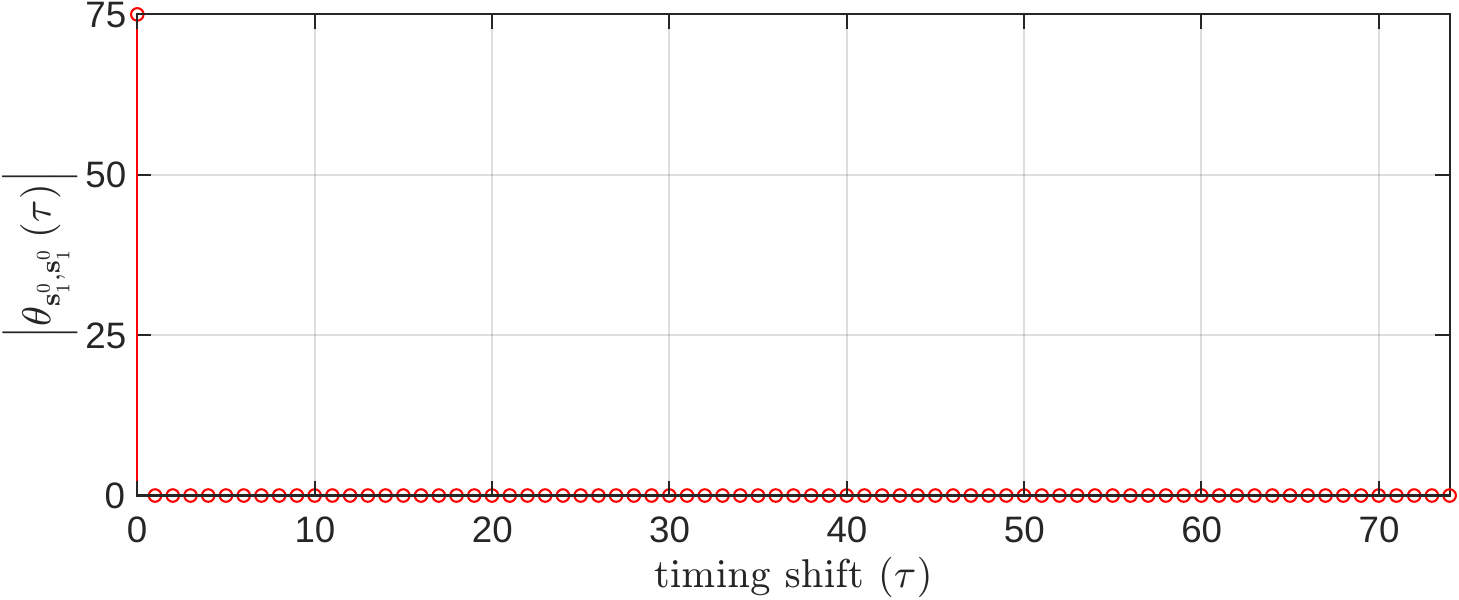}
        \caption{The PACF of ${\textbf{s}}_1^0$.}
        \label{A4}
    \end{subfigure}
    \begin{subfigure}{\linewidth}
        \includegraphics[width=0.85\textwidth]{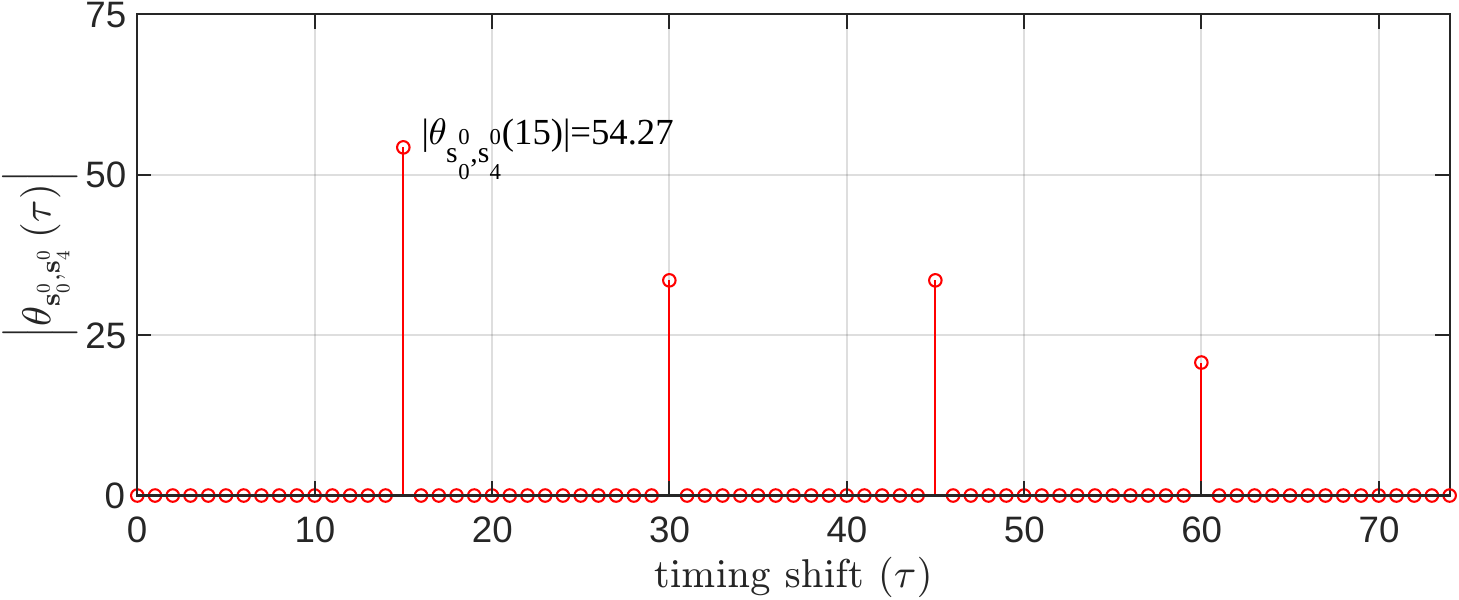}
        \caption{The PCCF of ${\textbf{s}}_0^0$ and ${\textbf{s}}_4^0$.}
        \label{A4}
    \end{subfigure}
    \begin{subfigure}{\linewidth}
        \includegraphics[width=0.85\textwidth]{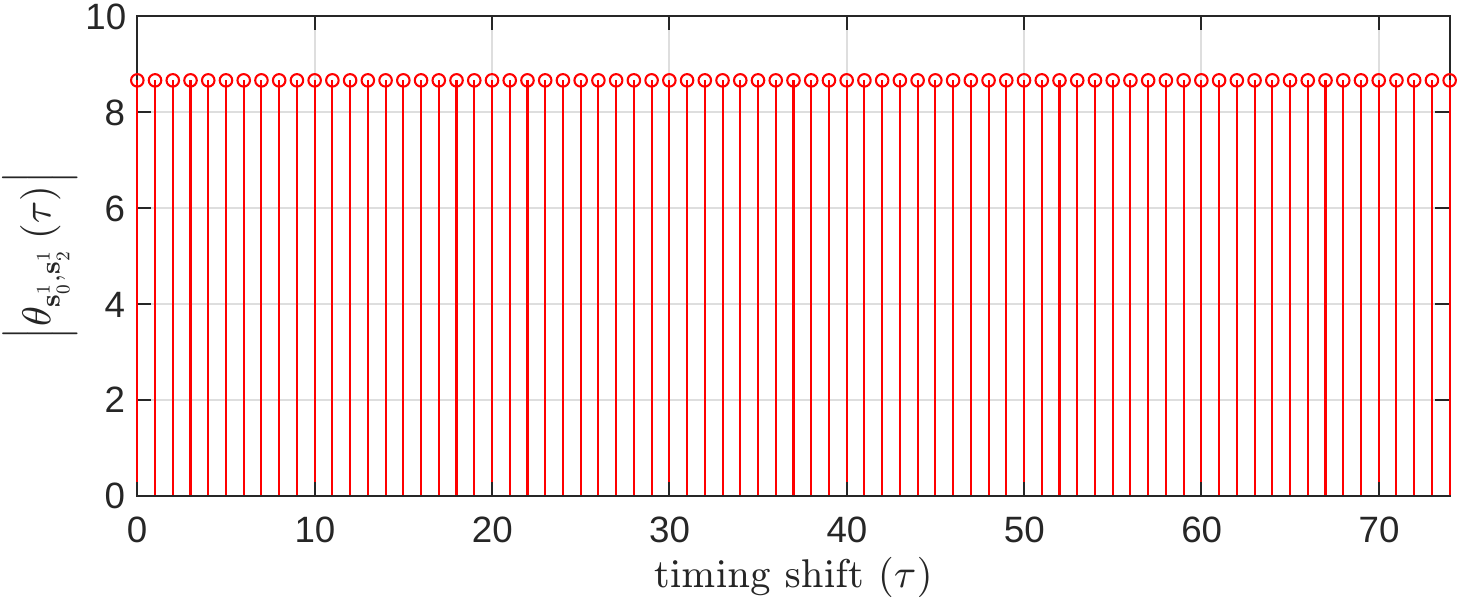}
        \caption{The PCCF of ${\textbf{s}}_1^0$ and ${\textbf{s}}_2^1$.}
        \label{A4}
    \end{subfigure}
    \caption{The correlation properties of Example 5.}
    \label{shiyan}
    \label{amvi}
\end{figure}

\textit{Theorem 3:} Let $R$ be even, $L=RT$, $N=RT^2$ and $M=1$. Select $M$ rows randomly from the circular Florentine array $F^q\left(0< q<\left(T-2 \right) ! \right)$ to form the index matrix $A$, where $F^q$ is obtained by Construction \uppercase\expandafter{\romannumeral1}. The ${P_u}\left( l\right)$ within the phase matrix $P$, associated with the sequence set $S$, is defined as:
\[{P_u}\left( {l} \right)={P_u}\left( {t + {r}T} \right) = w_{2R}^{{r}^2}w_T^{ut},\]
where $0 \le {l} < L$, $0 \le {r} < R$ and $0 \le t,u < T$.

According to the Main Framework, ${s_u\left( n \right)}$ in $S$ is defined as
\begin{equation}
\begin{aligned}
&{s_u}\left( n \right)= {s_u}\left( {t + lT} \right)\\
 = &{L^{ - 1}}T\sqrt R \sum\limits_{{r} = 0}^{R-1} {w_{2R}^{{r}^2}w_T^{ut}w_L^{  l\left( {{A}\left( t \right) + {r}T} \right)}}.
\end{aligned}
\end{equation}
Then, the sequence set $S$ has the following properties:
\begin{enumerate}
\item{Each sequence in $S$ is unimodular and  perfect.}
\item{$S$ is an optimal $\left( {{RT^2},T,RT} \right)$-ZCZ sequence set.}
\end{enumerate}

\textit{Proof:}  The detailed proof of Theorem 3 is provided in the Appendix.


\textit{Corollary 3:} Let $F$ be the circular Florentine array described in Construction \uppercase\expandafter{\romannumeral1}. Select $F$ to from the index matrix $A$. Define a $T\times L$ matrix $Q$, the $l$-th element of the $u$-th row of the matrix $Q$ is expressed as 
\[{Q_u}\left( {l} \right)={Q_u}\left( {t + {r}T} \right) = w_{2R}^{{r}^2}w_T^{ut},\]
where $0 \le {r} < R$ and $0 \le t,u < T$. The phase matrix $P$ is obtained by swapping the $\left( {rT - 1}\right)$-th and $\left( {rT - 2}\right)$-th column of the matrix ${Q}$ for $0 < r \le R$. The sequence set $S$ generated through (23) exhibits identical characteristics to those outlined in Theorem 3.

\textit{Example 6:} Let $N = RT^2$, where $R=2$, $T=6$ and $M=1$. Then we choose the index matrix ${A}$ as ${A}= \left[ {\begin{array}{*{20}{c}}
0&1&2&3&5&4 \end{array}} \right]$. The phase matrix $P$ is shown in (24).
 \begin{figure*}[ht] 
 	\centering
 	\begin{equation}	
\begin{array}{l}
P = \left[ {\begin{array}{*{20}{c}}
{w_{12}^0}&{w_{12}^0}&{w_{12}^0}&{w_{12}^0}&{w_{12}^0}&{w_{12}^0}&{w_{12}^3}&{w_{12}^3}&{w_{12}^3}&{w_{12}^3}&{w_{12}^3}&{w_{12}^3}\\
{w_{12}^0}&{w_{12}^2}&{w_{12}^4}&{w_{12}^6}&{w_{12}^8}&{w_{12}^{10}}&{w_{12}^3}&{w_{12}^5}&{w_{12}^7}&{w_{12}^9}&{w_{12}^{11}}&{w_{12}^1}\\
{w_{12}^0}&{w_{12}^4}&{w_{12}^8}&{w_{12}^0}&{w_{12}^4}&{w_{12}^8}&{w_{12}^3}&{w_{12}^7}&{w_{12}^{11}}&{w_{12}^3}&{w_{12}^7}&{w_{12}^{11}}\\
{w_{12}^0}&{w_{12}^6}&{w_{12}^0}&{w_{12}^6}&{w_{12}^0}&{w_{12}^6}&{w_{12}^3}&{w_{12}^9}&{w_{12}^3}&{w_{12}^9}&{w_{12}^3}&{w_{12}^9}\\
{w_{12}^0}&{w_{12}^8}&{w_{12}^4}&{w_{12}^0}&{w_{12}^8}&{w_{12}^4}&{w_{12}^3}&{w_{12}^{11}}&{w_{12}^7}&{w_{12}^3}&{w_{12}^{11}}&{w_{12}^7}\\
{w_{12}^0}&{w_{12}^{10}}&{w_{12}^8}&{w_{12}^6}&{w_{12}^4}&{w_{12}^{2}}&{w_{12}^3}&{w_{12}^{1}}&{w_{12}^{11}}&{w_{12}^9}&{w_{12}^7}&{w_{12}^5}
\end{array}} \right]
\end{array}
 \end{equation}
 \end{figure*}
From (23), a sequence set $S$ of size $6$ with period $72$ is obtained as
\begin{equation*}
\begin{aligned}
{s_u}\left( n \right) = {s_u}\left( {t + 6l} \right)= \frac{{\sqrt 2 }}{2}\sum\limits_{{r} = 0}^1 {{P_u}\left( {t + 6{r}} \right)w_{12}^{  l\left( {{A}\left( t \right) + 6{r}} \right)}}, 
\end{aligned}
\end{equation*} 
where $0 \le u,t < 6, 0 \le l<12$. The PACF of ${\textbf{s}}_4$ and PCCF between ${\textbf{s}}_1$ and ${\textbf{s}}_5$ are shown in Fig. 3. 

\begin{figure}[!h]
    \centering
    \begin{subfigure}{0.48\linewidth}
        \includegraphics[width=\textwidth]{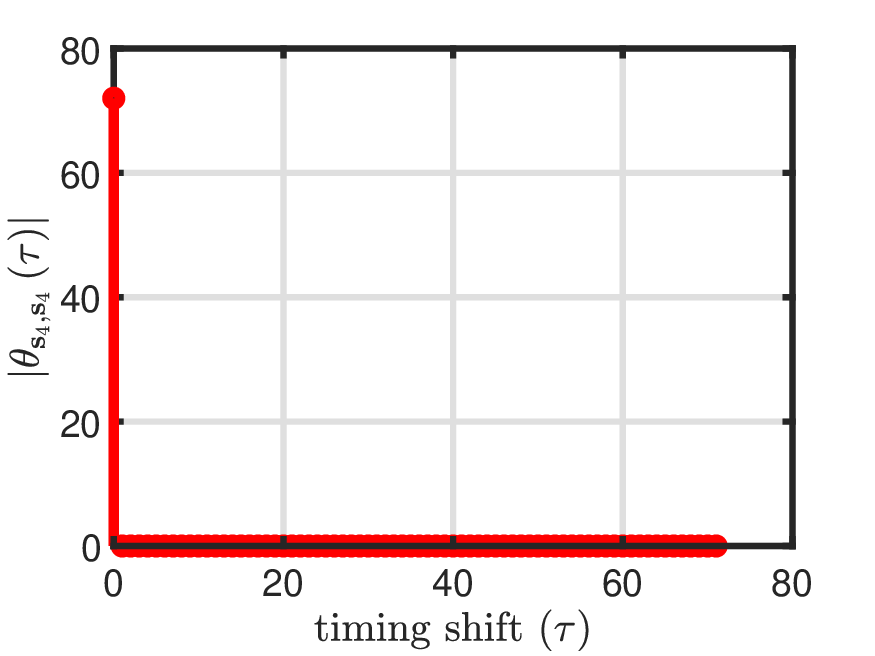}
        \caption{The PACF of ${\textbf{s}}_4$.}
        \label{A4}
    \end{subfigure}
    \begin{subfigure}{0.48\linewidth}
        \includegraphics[width=\textwidth]{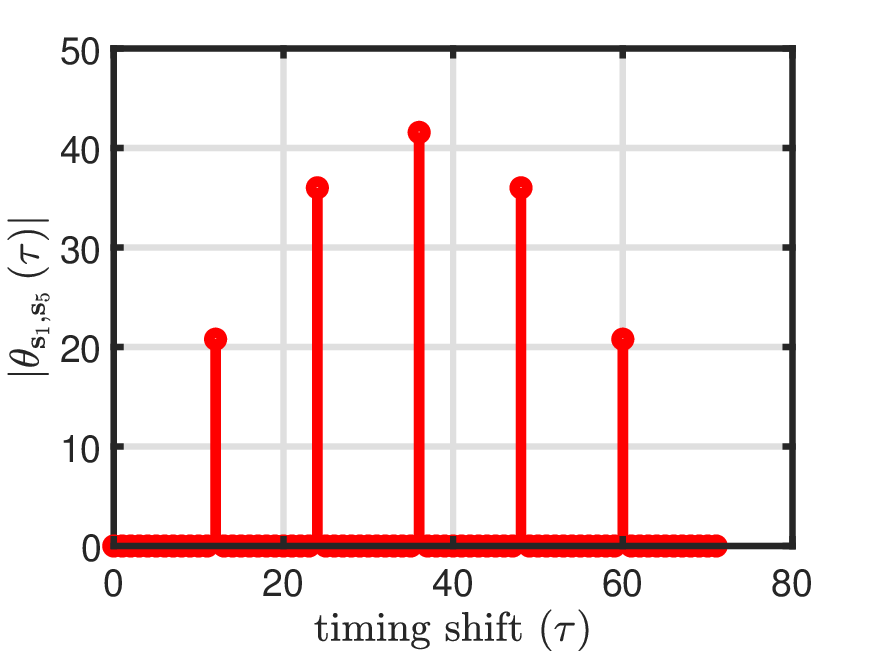}
        \caption{The PCCF of ${\textbf{s}}_1$ and ${\textbf{s}}_5$.}
        \label{A4}
    \end{subfigure}    
    \caption{The correlation properties of Example 6.}
    \label{shiyan}
    \label{amvi}
\end{figure}

\subsection{Comparison With The Previous Related Constructions}
In Table I, we list some known constructions of multiple ZCZ sequence sets with optimal correlation properties. From the construction perspective, existing methods in \cite{Popovic10}, \cite{Zhou17}, \cite{Zhang20} and \cite{Pitaval21} are developed from a time-domain approach. In contrast, our proposed method uniquely leverages the Zak transform, operating directly in the Zak domain. 

With regard to the cyclic equivalence and availability of sequences, existing methods suffer from several different limitations. Specifically, the Hadamard matrix based approach in \cite{Popovic10} may produce cyclically equivalent sequences. While the constructions in \cite{Zhou17} and \cite{Zhang20} can ensure unique sequences, the achievable sequence set numbers are limited due to their reliance on specific functions. Moreover, the number of multiple ZCZ sequence sets in \cite{Popovic10}, \cite{Zhou17}, \cite{Zhang20} and \cite{Pitaval21} is $d-1$, where $d$ is the smallest prime divisor of the sequence length. 

To address these limitations, \cite{Zhang22} utilizes circular Florentine arrays to create multiple optimal ZCZ sets with period $RT^2$, where $R$ is a positive integer. When $T$ is a prime, the number of the ZCZ sequence sets is also equal to $d-1$ as in \cite{Popovic10}, \cite{Zhou17} and \cite{Zhang20}. When $T$ is non-prime, the number of the ZCZ sequence sets depends on the number of rows of the cyclic Florentine array ${F_c}\left( T \right)$, which is strictly larger than $d-1$. Although \cite{Zhang22} leads to relatively large number of ZCZ sequence sets, their method still introduces cyclic equivalence within individual sets regardless of whether $R=1$ or not. 

{Note that cyclically distinct sequences are highly desirable in practice \cite{Zhou17}, \cite{Golomb2005}. On the contrary, the use of cyclically equivalent sequences could enable an attacker to easily decode the sequences of multiple users once the sequence of one user is decoded. This is unacceptable for secure information transmission in, for example, military and satellite communication systems.

Compared with \cite{Zhang22}, our proposed construction effectively avoids cyclic equivalence within individual sets. One can show that there are $(T-2)!$ distinct cases for the proposed sequence sets, thus offering a wide range of possibilities for various applications. On the other hand, the construction method in \cite{Song21} can only yield a single sequence set with optimal cross-correlation. Additionally, the application of cyclic Florentine array in the Zak-domain has not been reported before, to the best of our knowledge.}

\begin{table*}
\caption{The parameters for several sets of multiple ZCZ sequences}
\label{table}
\small
\centering
\setlength{\tabcolsep}{4pt}
\begin{tabular}{|m{1.0cm}<{\centering}|m{0.8cm}<{\centering}|m{0.8cm}<{\centering}|m{1.0cm}<{\centering}|m{0.5cm}<{\centering}|m{3.0cm}<{\centering}| m{1.0cm}<{\centering}|m{2.3cm}<{\centering}|m{2cm}<{\centering}|m{3cm}<{\centering}|}
\hline
Methods & Period& Phase Number &  Set size & $H_{z}$ & The number of $ZCZ$ sets & $\theta_c\left( {{\cal S}} \right)$ & Whether cyclically distinct & Number of distinct sequence sets & Note\\
\hline
\cite{Popovic10} &$RT^2$ & $RT$ &  $T$ & $RT$ & $d-1$ & $\sqrt R T $&N&1&$R$ and $T$ are positive integers.\\
\hline
\cite{Zhou17} &$T^2$ &$T$ &  $T$ & $T$ & $T-1$ & $T$&Y&1&$T$ is odd prime.\\
\hline
\cite{Zhang20}&$T^2$ &$T$ &  $T$ & $T$ & $d-1$ & $T$ &Y&1& $T$ is odd.\\

\hline
 \cite{Pitaval21} & $T$ &$T$ & $ \left\lfloor {\frac{T}{L}} \right\rfloor $&$L$  &$T-1$&$\sqrt T $   &N&1&$T$ is prime.\\
 \hline
 \multirow{3}{*}{\cite{Zhang22}} &$RT^2$ &$RT$ &  $T$ &$T$&  $d-1$ & $\sqrt R T$&N&1&$T$ is prime. $R$ is a positive integer. \\
  
 \cline{2-9}
 &$T^2$ &$T$ &  $T$ &$T$&  $F_c(T)$ & $T$&N&1 &$T$ is nonprime.\\
 \cline{2-9}
 &$RT^2$ &$RT$ &  $T$ &$RT$&  $min\{R^*-1,F_c(T)\}$ & $\sqrt R T$&N&1&$T$ is nonprime and $R\neq 1$ is a positive integer.\\
 \hline
Th.1 & $T^2$ &$T$ &$T$ & $T$ & $F_c(T)$ & $T$&Y&$\left(T-2 \right)!$&$T>3$ is a integer.\\
\hline
Th.2 & $RT^2$ &$RT$ &$T$ & $RT$ & $min\{R^*-1,F_c(T)\}$ &$\sqrt R T$&Y&$\left(T-2 \right)!$&$R$ is odd, $T>3$ is a integer.\\
\hline
Th.3 & $RT^2$ &$2RT$ &$T$ & $RT$ & $1$ &$-$&Y&$\left(T-2 \right)!$&$R$ is even, $T>3$ is a integer.\\
\hline
\end{tabular}
\begin{tablenotes} 
\footnotesize 
\item \noindent{ $d$ is the smallest prime divisor of the period; $R^*$ is the smallest prime divisor of $R$.}
\end{tablenotes}   
\label{tab1}
\end{table*}

\subsection{Evaluation of the Synchronization performance in OTFS}

{Let us consider a wireless communication system where $L \times T$ data symbols in the DD domain are modulated using OTFS over a total bandwidth $B$ operating at the carrier frequency $f_c$, where $L$ and $T$ denote the numbers of delay bins and Doppler bins, respectively. Firstly, the OTFS modulator distributes these symbols $\left\{ {{X(j,t)},j{\rm{ }} = 0, \ldots ,{\rm{ }}L - 1,{\rm{ }}t{\rm{ }} = {\rm{ }}0, \ldots ,T - 1} \right\}$ in the two-dimensional DD grid. Denote by $\Delta f = B/T$ the frequency spacing and $\Gamma  = 1/\Delta f$ the corresponding time duration. Thus, the duration of one OTFS frame is $L\Gamma$.

The DD-domain symbols $X(j,t)$ are then transformed into the TF domain $\left\{ {X_{\mathrm{TF}}}(l,m), {l = 0,\ldots,L - 1},{m = 0, \ldots ,T - 1} \right\}$ via ISFFT, as shown in (25) below. 
\begin{equation}
{X_{\rm{TF}}}(l,m) = \frac{1}{{\sqrt {LT} }}\sum\limits_{j = 0}^{L - 1} {\sum\limits_{t = 0}^{T - 1} {X(j,t)} {e^{i2\pi (\frac{{lj}}{L} - \frac{{mt}}{T})}}}.
\label{OTFS1}
\end{equation}

After mapping into TF domain, Heisenberg transform is applied to generate the discrete time-domain signal as follows:
\begin{equation}
s({t} + {l}T) = \frac{1}{{\sqrt {T} }}\sum\limits_{m = 0}^{T - 1} {{X_{{\rm{TF}}}}} ({l},m){e^{\frac{{2\pi im}}{T}{t}}}.
\label{OTFS2} 
\end{equation}

Substituting (\ref{OTFS1}) into (\ref{OTFS2}), we obtain 
\begin{equation*}
s(t + lT) = \frac{1}{{\sqrt L }}\sum\limits_{j = 0}^{L - 1} {{X(j,t)} {e^{i2\pi \frac{l}{L}j}}},
\label{OTFS3}
\end{equation*}
which is equivalent to IFZT defined in (4). The relationship of OTFS modulation and IFZT is shown in Fig. \ref{figzak3}. 
\begin{figure}[h]
\centering
\includegraphics[width=2.5in]{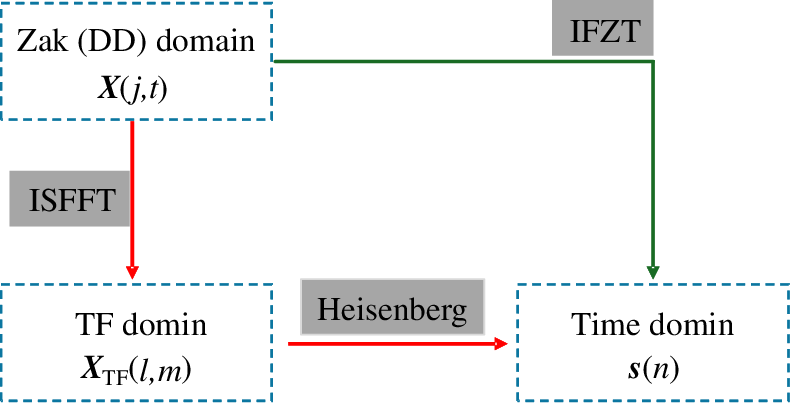}
\caption{The relationship of OTFS modulation and IFZT.}
\label{figzak3}
\end{figure}

Furthermore, let us consider a doubly selective channel consisting of $C$ paths as shown below:  
\begin{equation}
h(\tau ,v) = \sum\limits_{\rho = 1}^C {{h_p}\delta (\tau  - {\tau _\rho})} \delta (v - {v_\rho}),
\end{equation}
where $h_\rho$, $\tau _\rho$ and $v_\rho$ represent the channel fading coefficient, delay and Doppler values of the $\rho $-th path, respectively. To determine $L$ and $T$, it is required that $\max (\tau ) < \Gamma$ and $\max (v) <  \Delta f$.
After the OTFS modulation (i.e., IFZT transform), the time-domain signal is represented by $s(n)$. The received signal $r(n)$ can then be expressed as 
\begin{equation}
r(n) = \int{\int {h(\tau ,v)s(n - \tau ){e^{2\pi iv(n - \tau )}}d\tau dv} +{\cal N}},
\end{equation}
where ${\cal N}$ denotes the additive white Gaussian noise term with variance of $\sigma^2$.

To understand the Doppler resilience of the proposed Zak sequences, we also introduce the definition of discrete periodic auto-ambiguity function (auto-AF) of sequences as follows\cite{Tian2025}:

Let ${\textbf{s}}=\left( {{s\left( 0 \right)},...,{s\left( N-1 \right)}} \right)$ be a sequence of period $N$. The periodic auto-AF of ${\textbf{s}}$ at time shift $\tau$ and Doppler shift $v$ is defined as 
\begin{equation}
{{AF_{\textbf{s}}}\left( \tau,v  \right) = \sum\limits_{n = 0}^{N - 1} {s\left( n+\tau\right)_N{s^*}{{\left( n  \right)}}w_{N}^{vn}},} 
\end{equation}
where $-N < \tau, v  < N$.



Next, we investigate the synchronization performance of a single-input single-output (SISO) OTFS system. Our idea is to transmit a sparse Zak matrix (satisfying the perfect sequence condition as specified in Subsection IV.A) as a preamble sequence (frame) in the DD domain and then leverage its zero auto-correlation sidelobes for synchronization in the time domain. Fig. \ref{figzak2} illustrates the synchronization model for such a SISO-OTFS system. We assume that a cyclic prefix (CP) is added to the beginning of each frame for mitigation of inter-frame interference. 
The length of the sliding window is $L\Gamma$. The receive signal in the window is correlated with the known reference sequence (i.e., the time-domain sequence of the aforementioned sparse Zak matrix) each time within the timing acquisition range $T_u$ to detect the starting position of the preamble sequence. The synchronization point is detected once a correlation peak is achieved at certain time shift. 
\begin{figure}[h]
\centering
\includegraphics[width=3.5in]{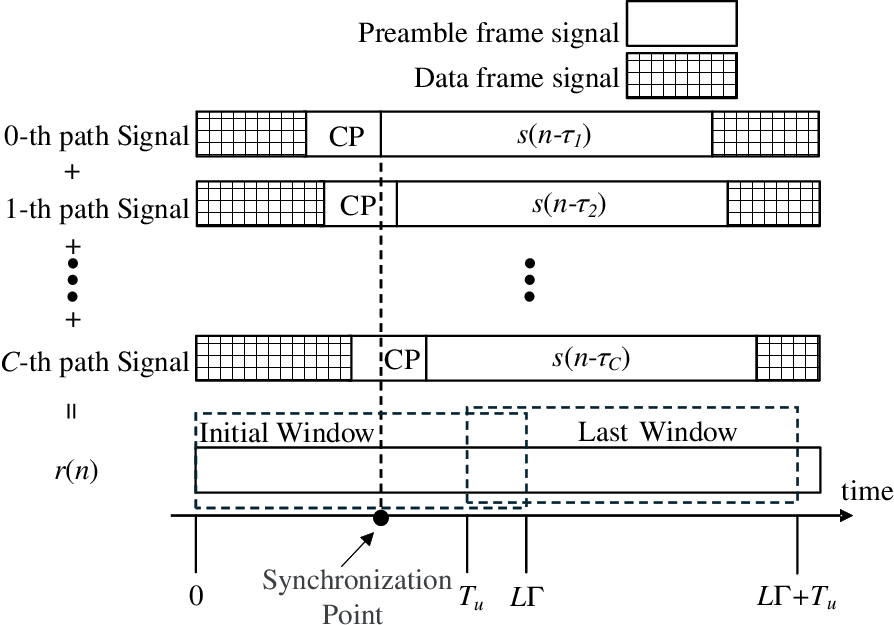}
\caption{Synchronization model of SISO-OTFS system.}
\label{figzak2}
\end{figure}

\begin{align}
\label{X168}
\notag & X_1^0 = \\
&8 \sqrt{2}\left[ {\begin{array}{*{20}{l}}
{w_{16}^0}&0&0&0&0&0&0&0\\
0&{w_{16}^2}&0&0&0&0&0&0\\
0&0&0&0&0&0&{w_{16}^{12}}&0\\
0&0&{w_{16}^4}&0&0&0&0&0\\
0&0&0&0&0&{w_{16}^{10}}&0&0\\
0&0&0&{w_{16}^6}&0&0&0&0\\
0&0&0&0&0&0&0&{w_{16}^{14}}\\
0&0&0&0&{w_{16}^8}&0&0&0\\
{w_{16}^4}&0&0&0&0&0&0&0\\
0&{w_{16}^6}&0&0&0&0&0&0\\
0&0&0&0&0&0&{w_{16}^0}&0\\
0&0&{w_{16}^8}&0&0&0&0&0\\
0&0&0&0&0&{w_{16}^{14}}&0&0\\
0&0&0&{w_{16}^{10}}&0&0&0&0\\
0&0&0&0&0&0&0&{w_{16}^2}\\
0&0&0&0&{w_{16}^{12}}&0&0&0
\end{array}} \right]
\end{align}
\begin{figure}[!h]
    \centering
    \includegraphics[width=0.48\textwidth]{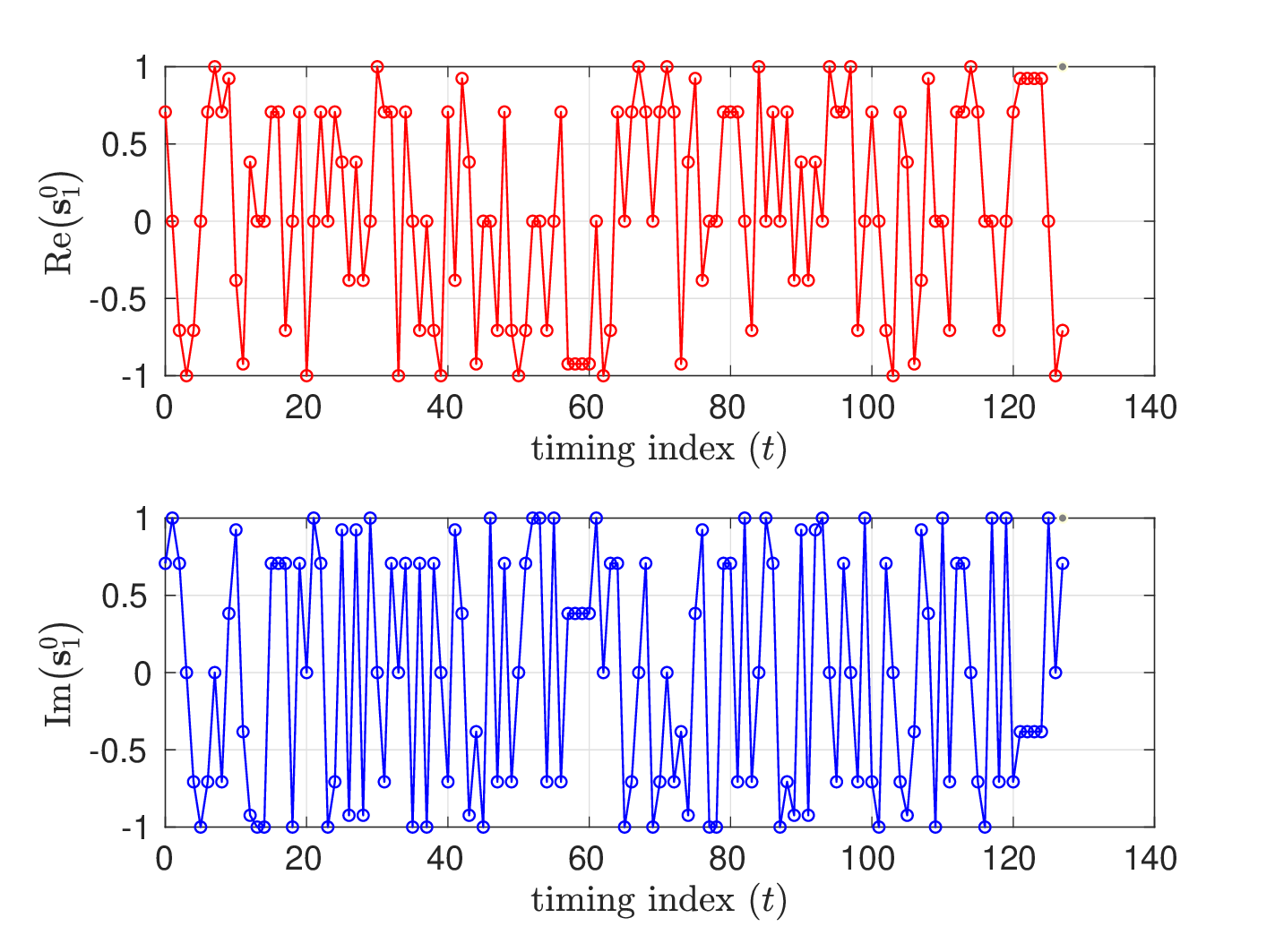}
    \caption{The real and imaginary parts of $\mathbf{s}_1^0$}.
    \label{realimag}
\end{figure}
\begin{figure}[!h]
    \centering
    \begin{subfigure}{0.45\textwidth}
        \includegraphics[width=\textwidth]{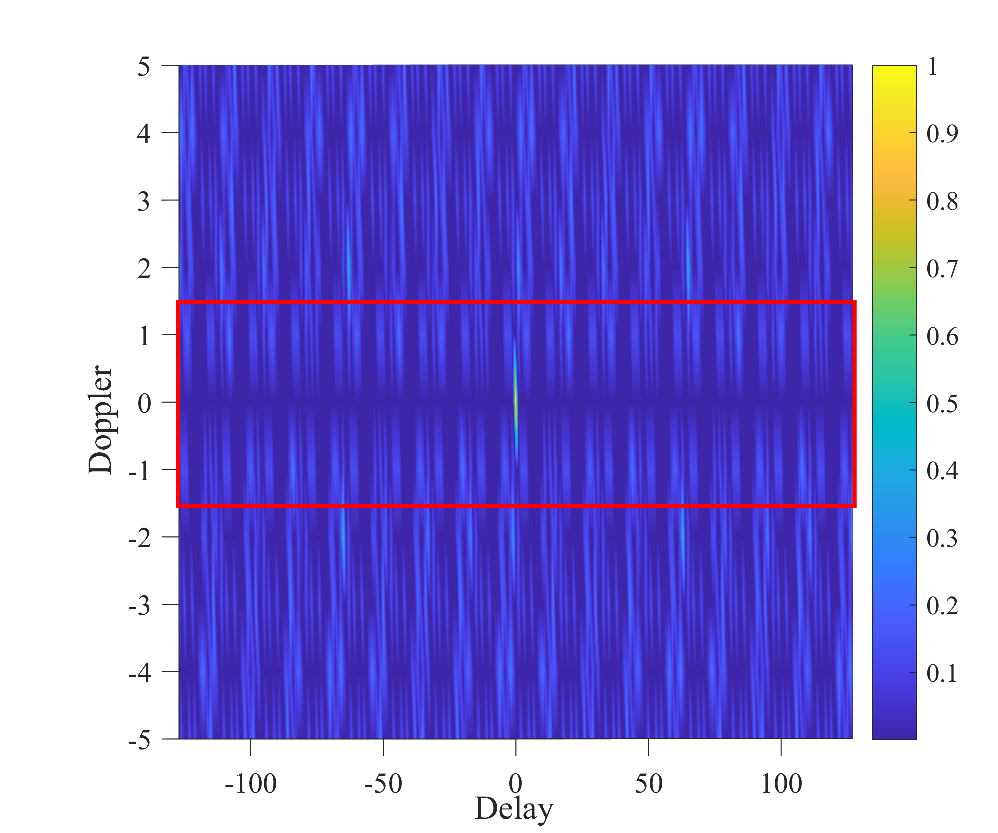}
        \caption{The periodic auto-AF of $\mathbf{s}_1^0$.}
        \label{A4}
    \end{subfigure}
    \begin{subfigure}{0.45\textwidth}
        \includegraphics[width=\textwidth]{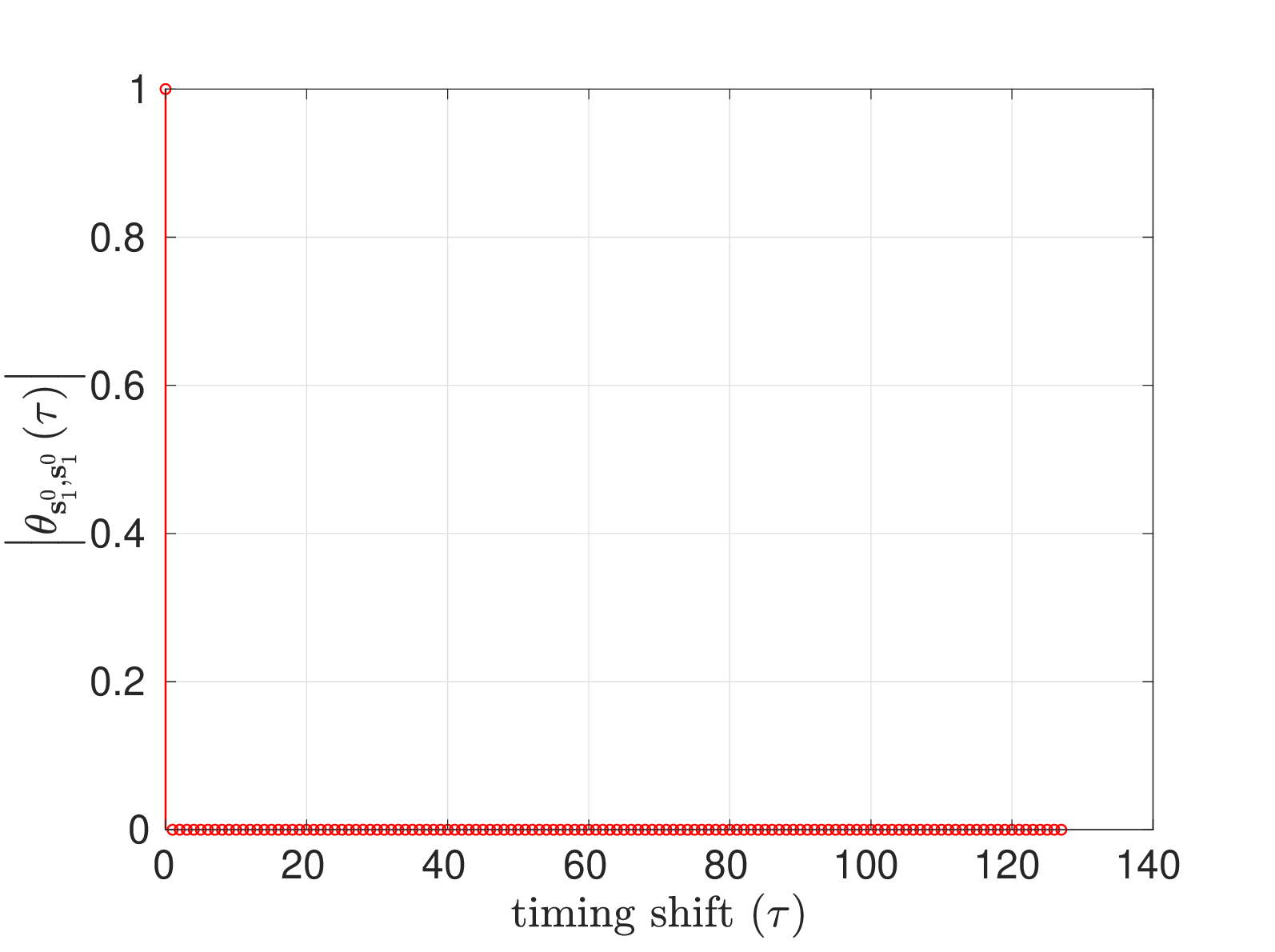}
        \caption{The PACF of $\mathbf{s}_1^0$}.
        \label{A4}
    \end{subfigure}
    \caption{The periodic auto-AF and PACF of $\mathbf{s}_1^0$}
    \label{shiyan}
    \label{amvi}
\end{figure}

For evaluation, three consecutive $16\times8$ OTFS frames are transmitted, with the second frame as the synchronization frame and the remaining frames carrying QPSK data in the DD domain. Equal transmission power is assumed for all frames. As an example, we consider Zak matrix $X_1^0$ in (\ref{X168}) which is constructed via Theorem 3 with parameters $R=2$ and $T=8$. The corresponding time-domain sequence is $\mathbf{s}_1^0$. The real and imaginary parts of $\mathbf{s}_1^0$ are shown in Fig. \ref{realimag}. Also, as shown in Fig. \ref{amvi}, such a sequence exhibits excellent periodic auto-AF with strong resilience to Doppler as well as perfect periodic auto-correlation sidelobes.  

To introduce randomness for the starting point of the synchronization frame, some initial portion of the first data frame is randomly truncated. {For numerical simulation, we consider $C\in \{3, 6\}$. The channel coefficients $h_p$ are generated by $h_p\sim \mathcal{C} \mathcal{N} (0,q_p)$ \cite{Bultitude2007}, where $q_p=\exp \left( -\tau _p\frac{r_{\tau}-1}{r_{\tau}\sigma _{\tau}} \right) \cdot 10^{\frac{-Z_p}{10}}$, $r_{\tau}$ is the proportionality factor, $\sigma _{\tau}$ is the root mean square (RMS) delay spread, $Z_p$ is additional fading. The normalized delay of $C$ paths is $\left[ 0,1,\cdots ,C-1 \right] $. Each normalized delay has a single Doppler shift generated using Jakes' formula $\nu _p=\nu _{\max}\cos\mathrm{(}\theta _p)$, where $\nu _{\max}$ is the maximum Doppler shift determined by the maximum relative velocity $v_{\max}$, $\theta _p$ is uniformly distributed over $\left[0,2\pi\right]$.} The system parameters are summarized in Table \ref{tab1}. The synchronization success probability is evaluated using $10000$ Monte Carlo simulations at each signal-to-noise ratio (SNR). 
\begin{table}[!h]
	\begin{center}
		\caption{SYSTEM PARAMETERS}
		\label{tab1}
		\begin{tabular}{| c | c | }
			\hline
			number of Doppler bins ($T$) & 8 \\
			\hline
			number of delay bins ($L$) &16\\
			\hline
			Carrier Frequency ($f_c$) &$6$ GHz \\ 
			\hline
			frequency spacing ($\Delta f$) & 15 KHz  \\
			\hline
			Number of paths ($C$)  & 3, 6  \\
			\hline
			maximum relative velocity ($v_{\max}$)  & 200 km/h  \\
			\hline
			proportionality factor ($r_\tau$)  & 2.3  \\
			\hline
			RMS delay spread ($\sigma_{\tau}$)  & $1.5$ $\text{\textmu}$s  \\ 
			\hline
			additional fading ($Z_p$)     & $0$ dB  \\ 
			
			\hline 
			CP Length & 32 \\
			\hline 
			Sliding Window Length & 128 \\
			\hline 
		\end{tabular}
	\end{center}
\end{table}

{Fig. \ref{v_syn} compares the synchronization probabilities under different maximum velocities at $\text{SNR}=20$ dB. For $C\in \{3,6\}$, our simulation results indicate that increasing $v_{\max}$ leads to a deterioration in synchronization performance. In our synchronization algorithm, to obtain a distinct correlation peak, the Doppler shift of the main path is estimated and compensated. Since the remaining receive signal paths are regarded as noise, the synchronization performance of the proposed Zak sequence cannot reach $100\%$ at higher $v_{\max}$. As a result, an error floor appears in the subsequent bit error rate (BER) simulation results. At the same time, compared with random Zak sequences, the proposed Zak sequences achieve significantly improved synchronization performance, thanks to its excellent autocorrelation and ambiguity properties.}

Fig.~\ref{sny} evaluates the synchronization success probability performance as a function of SNR. In fact, due to the noise, the multipath propagation, as well as the sequence autocorrelation sidelobes, the receiver may fail to achieve synchronization, thus potentially leading to a catastrophic degradation of decoding performance. Thanks to the perfect auto-correlation property of the proposed sequence $\mathbf{s}_1^0$, the receiver can effectively detect the starting point of the preamble sequence through sliding window correlation. Our proposed preamble sequence can achieve a success rate close to $100\%$ at SNR of 10 dB or higher, while the random sequence (with random QPSK symbols in the DD domain) can only reach about $89\%$ for the 6-path case at SNR of 20 dB or higher. {In addition, due to the perfect auto-correlation property of the proposed Zak sequence, it exhibits robust synchronization performance for different numbers of paths.} 

{Fig. \ref{BER} shows the BER simulation results under different SNRs after synchronization, by assuming that perfect channel fading coefficients are known. At high SNR, since the synchronization performance of the proposed Zak sequence approaches but cannot reach $100\%$, any occasional synchronization failure results in a substantial gap between its BER curve and that with perfect synchronization. In addition, the BER performances of the proposed Zak sequences are significantly better than that of random Zak sequences, especially for the case of $C=6$.}

\begin{figure}[!h]
	\centering
	\includegraphics[width=0.9\linewidth]{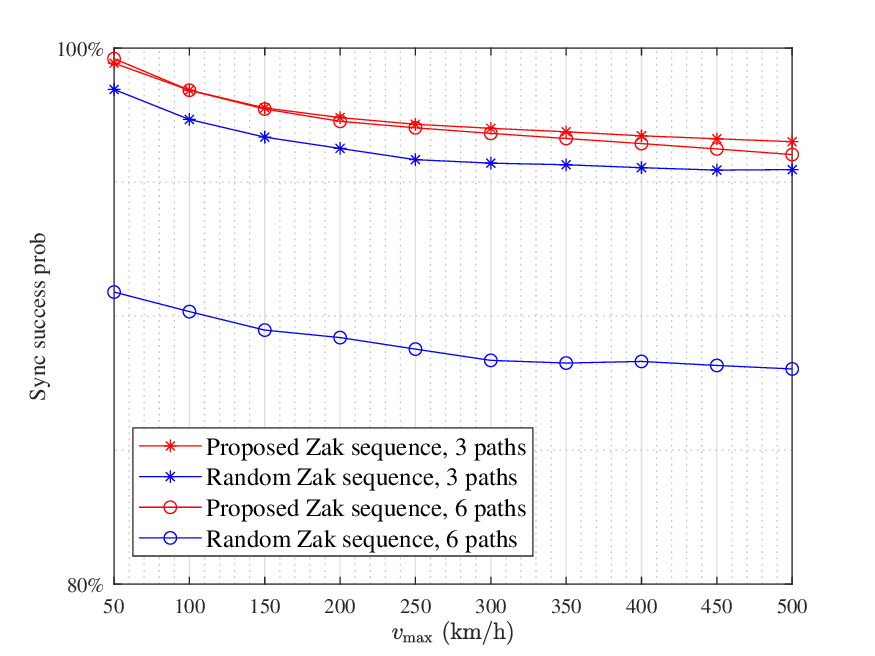}
	\caption{{Comparison of synchronization success probabilities under different $v_{\max}$.}}
	\label{v_syn}
\end{figure}
\begin{figure}[!h]
    \centering
    \includegraphics[width=0.9\linewidth]{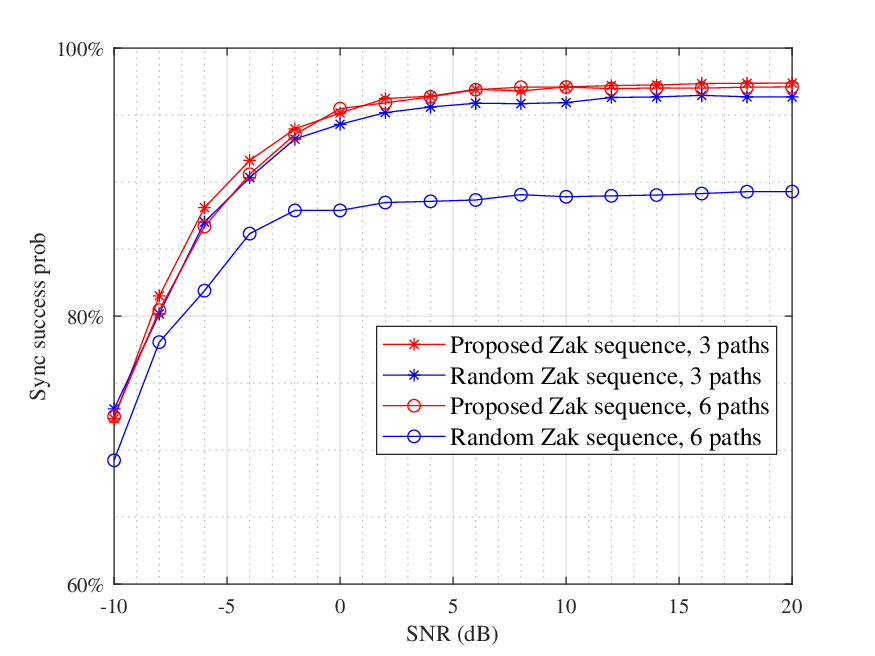}
   \caption{{The synchronization success probability versus SNR.}}
    \label{sny}
\end{figure}
\begin{figure}[!h]
    \centering
    \includegraphics[width=0.9\linewidth]{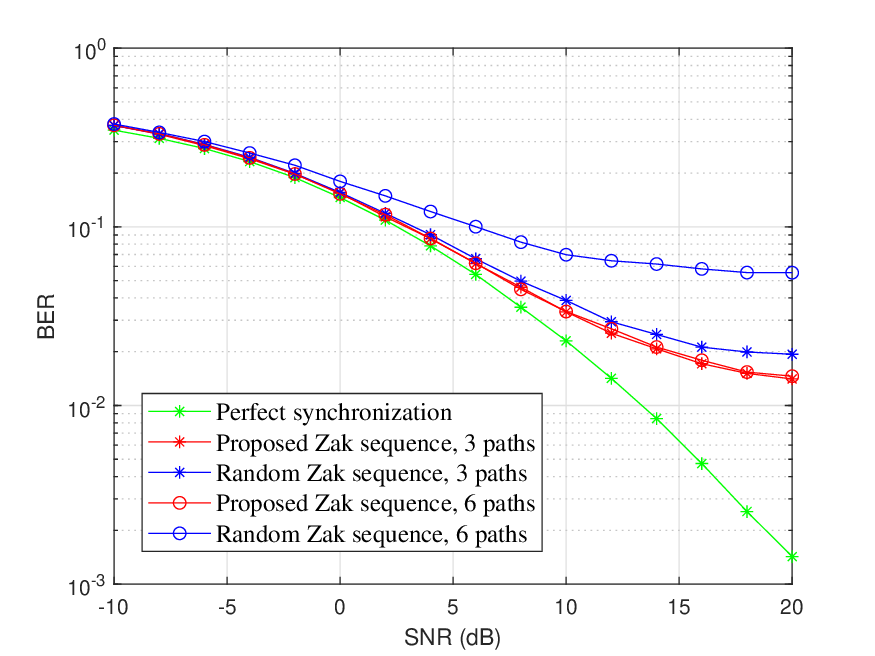}
     \caption{{BER results versus SNR with perfect channel fading coefficients and by using the proposed synchronization algorithm.}}
    \label{BER}
\end{figure}

}
\section{Conclusions}
In this paper, we have presented a novel framework for constructing multiple ZCZ sequence sets with optimal correlation properties using IFZT. To ensure sequence sparsity in the Zak domain, we have introduced index matrices and phase matrices that are associated to FZT. The admissible conditions of these matrices are also derived. It has been found that the maximum inter-set cross-correlation can beat the Sarwate bound provided that a circular Florentine array is adopted as the index matrix. {Besides, we have demonstrated that the Zak-domain-induced optimal sequences can be employed as preamble sequences in the DD domain for excellent OTFS synchronization performance. 

As a future work, it is interesting to analyze and study the ambiguity properties of the proposed sequences \cite{LAZ2022}. Moreover, it is worthy to investigate their applications for channel estimation and sensing in, for example, multi-user MIMO-OTFS systems. To this end, one may leverage the multiple ZCZ set properties as well as the excellent ambiguity sidelobes of these sequences. For spectrum-efficient transmission, one may also superimpose those sparse Zak matrices in the DD domain with random communication data symbols. The readers are invited to attack these research problems.}

{\appendices
\section*{Proof of the Theorem 1}
\textit{Proof:} Initially demonstrate that the first property of the sequence sets is met. For $R=1$, $0 \le t<T$ and $0 \le l<L$, we have
\[\left| {\sum\limits_{{r} = 0}^{R - 1} {{P_u}\left( {t + {r}T} \right)w_L^{  l{r}T}} } \right|  =\left| {P_u}\left( t \right)\right|=1.\]

Additionally, $A$ is a circular Florentine array, it is known that $A^m$ is a permutation on $\mathbb{Z}_T$. And we have
 \[\left| {P_u\left( {t+ {r}T} \right)} \right| = \left| {P_u\left( t \right)} \right| = 1.\] 
According to Lemma 5 and 6, any sequence in ${S^m}$ obtained by Theorem 1 is unimodular and perfect.  This completes the proof of Part 1).

We now prove Part 2). Let ${\textbf{s}_u^m}$ and ${\textbf{s}_v^m}$ be two sequences
in $S^m$, where $0 \le u\ne v < T$ and $0 \le m <M$. Based on the IFZT and Lemma 7, for $\tau_1=0$, we distinguish between the following two cases to calculate (15).

Case 1: When ${\tau_2} = 0$, (15) becomes
\begin{equation*}
\begin{aligned}
& {\sum\limits_{{r} = 0}^{R - 1} {\sum\limits_{t = 0}^{T - 1} {w_L^{  0\left( {A^m\left( t\right) + {r}T} \right)}{P_u}\left( {t + {r}T} \right)P_v^{*}\left( {t + {r}T} \right)} }}\\
 =& \sum\limits_{t = 0}^{T - 1} {{P_u}\left( t\right)P_v^{*}\left( t \right)} = \sum\limits_{t = 0}^{T - 1} {w_T^{\left( {u - v} \right)t}}.
\end{aligned}
\end{equation*}
The result follows from the fact that $\left( {{u - v}} \right)t$ is a permutation on  $\mathbb{Z}_T$ for any $u \ne v$.

Case 2: When ${\tau_2} =1$, (15) becomes
\begin{equation*}
\begin{aligned}
& {\sum\limits_{{r} = 0}^{R - 1} {\sum\limits_{t = 0}^{T - 1} {w_T^{  \left( {A^m\left( t\right) + {r}T} \right)}{P_u}\left( {t + {r}T} \right)P_v^{*}\left( {t + {r}T} \right)} } } \\
 =& \sum\limits_{t = 0}^{T - 1} {w_T^{  A^m\left( t\right)}{P_u}\left( t \right)P_v^{*}\left( t \right)}=\sum\limits_{{t} = 0}^{T -1} {w_T^{  A^m\left( {{t}} \right) + \left( {u - v} \right){t}}}.
\end{aligned}
\end{equation*}

Within Construction \uppercase\expandafter{\romannumeral1}, it is established that $A^m\left( {{t}} \right) \ne at$, where $a$ is a constant. This implies that $ A^m\left( {{t}} \right) + \left( {u - v} \right)t$ is neither 0 nor a multiple of $t$. Then,  $0< \left| {\sum\limits_{{t} = 0}^{T -1} {w_T^{  A^m\left( {{t}} \right) + \left( {u - v} \right){t}}} } \right|<T$ and $0 < \left| {{\theta _{{\bf{s}}_u^m,{\bf{s}}_v^m}}\left( T  \right) } \right| < T^2$.

Hence, the sequence set $S^m$ is an optimal $\left( {{T^2},T,T} \right) $-ZCZ set for the Tang-Fan-Matsufuji bound. Furthermore, all sequences in ${S^m}$ exhibit cyclic distinctness.

We now proceed to the third part of the theorem demonstration. Let's consider any two sequences ${\textbf{s}_u^{m_1}}$ and ${\textbf{s}_v^{m_2}}$ within $S^{m_1}$ and $S^{m_2}$ respectively, where $0 \le u\ne v < T$ and $0 \le {m_1} \ne {m_2} <M$. Let $n = t + lT, \tau  = {\tau _1} + {\tau _2}T$, with $0 \le t, {\tau _1} < T$ and $0 \le l, {\tau _2} < L$. Based on the equation 
\begin{equation*}
\label{e:barwq}
\begin{split}
&\left| {\sum\limits_{{r} = 0}^{R - 1} {P_u^{{m_1}}\left( {t + {\tau _1}+{r}T} \right){P_v^{{m_2}*}\left( {t +{r}T} \right)w_R^{ {r}{\tau _2}}} }} \right|\\
= &\left|{P_u^{m_1}\left( {t+{\tau _1}  } \right)} {P_v^{{m_2}*}\left( {t } \right)} \right|,
\end{split}
\end{equation*}
since each element of the phase matrix is a power of $w_T$, this implies that
\[\left| {\sum\limits_{{r} = 0}^{R - 1} {P_u^{{m_1}}\left( {t + {\tau _1}+{r}T} \right){P_v^{{m_2}*}\left( {t +{r}T} \right)w_R^{ {r}{\tau _2}}} }} \right|=1.\]

Recalling that $A$ is an ${F_c}\left( T \right) \times T$ circular Florentine array over ${\mathbb{Z}_T}$, from Lemma 8, we can assert that ${\theta _{{\bf{s}}_u^{{m_1}},{\bf{s}}_v^{{m_2}}}}\left( \tau  \right) = T$ for any $0 \le {m_1} \ne {m_2} <M, 0 \le u\ne v < T$ and $0 \le \tau  < {T^2}$.
\hfill
$\hfill\blacksquare$ 
\section*{Proof of the Theorem 2}
\textit{Proof:} For $R$ being odd,  $0 \le t<T$ and $0 \le l<L$, we have

\begin{equation}
\begin{aligned}
&\left|{\sum\limits_{{r} = 0}^{R - 1} {{P_u^m}\left( {t + {r}T} \right)w_L^{  l{r}T}} } \right|^2\\
 =&\left( {\sum\limits_{{r} = 0}^{R - 1} {w_R^{\left( {m + 1} \right)\frac{{{r}\left( {1 + {r}} \right)}}{2}}w_T^{ut}w_L^{  l{r}T}} } \right)\\
& {\left( {\sum\limits_{r{'} = 0}^{R - 1} {w_R^{\left( {m + 1} \right)\frac{{r{'}\left( {1 + r{'}} \right)}}{2}}w_T^{ut}w_L^{  lr{'}T}} } \right)^*}\\
=&\left( {\sum\limits_{{r} = 0}^{R - 1} {w_R^{\left( {m + 1} \right)\frac{{{r}\left( {1 + {r}} \right)}}{2}}w_R^{  l{r}}} } \right)\\
&{\left( {\sum\limits_{r{'} = 0}^{R - 1} {w_R^{-\left( {m + 1} \right)\frac{{r{'}\left( {1 + r{'}} \right)}}{2}}w_R^{  -lr{'}}} } \right)}
\end{aligned}
\end{equation}

By analyzing the behavior of the summation $\sum\limits_{{r} = 0}^{R - 1} {w_R^{\left( {m + 1} \right)\frac{{{r}\left( {1 + {r}} \right)}}{2}}w_R^{  l{r}}}$ for various values of $l$ and $m$, where $0 \le l<L$ and $0 \le m<M$, (30) is reduced to
\begin{equation}
\begin{aligned}
&\left|{\sum\limits_{{r} = 0}^{R - 1} {{P_u^m}\left( {t + {r}T} \right)w_L^{  l{r}T}} } \right|^2\\
=&\sum\limits_{{r} = 0}^{R - 1} {w_R^{\frac{{{r}\left( {1 + r} \right)}}{2}}}  {\sum\limits_{{r'} = 0}^{R - 1} {w_R^{\frac{{-{r'}\left( {1 + r'} \right)}}{2}}}  }
\end{aligned}
\end{equation}

Denoting $a = \left\lfloor {\frac{R}{2}} \right\rfloor $, where $\left\lfloor {x} \right\rfloor $ represents the greatest integer less than or equal to $x$, we can reformulate equation (31) as follows:
\begin{equation*}
\begin{aligned}
&\sum\limits_{{r} = 0}^{R - 1} {w_R^{\frac{{{r}\left( {1 + r} \right)}}{2}}}  {\sum\limits_{{r'} = 0}^{R - 1} {w_R^{\frac{{-{r'}\left( {1 + r'} \right)}}{2}}}  }\\
=& \left( {2\sum\limits_{{r} = 0}^{a - 1} {w_R^{ {\frac{{{r}\left( {1 + {r}} \right)}}{2}}}}  + w_R^{ {\frac{{a\left( {1 + a} \right)}}{2}} }} \right)\\
&\left( {2\sum\limits_{{r'} = 0}^{a - 1} {w_R^{ {\frac{{-{r'}\left( {1 + {r'}} \right)}}{2}} }}  + w_R^{{\frac{-{a\left( {1 + a} \right)}}{2}} }} \right)\\
 = &4 {\sum\limits_{{r} = 0}^{a - 1} {w_R^{{\frac{{{r}\left( {1 + {r}} \right)}}{2}} }} }  {\sum\limits_{{r'} = 0}^{a - 1} {w_R^{  {\frac{{-{r'}\left( {1 + {r'}} \right)}}{2}} }} } +1 \\
&+ 2w_R^{ {\frac{{a\left( {1 + a} \right)}}{2}} }{\sum\limits_{{r'} = 0}^{a - 1} {w_R^{  {\frac{{-{r'}\left( {1 + {r'}} \right)}}{2}}}} } +2w_R^{ {\frac{-{a\left( {1 + a} \right)}}{2}}}  {\sum\limits_{{r} = 0}^{a - 1} {w_R^{ {\frac{{{r}\left( {1 + {r}} \right)}}{2}}}} }  \\
=& 4\left( {a - \frac{{{a^2} - a}}{{R - 1}}} \right) - 2\frac{{2a}}{{R - 1}} + 1\\
= &4a - \frac{{4{a^2}}}{{R - 1}} + 1=R.\\
\end{aligned}
\end{equation*}

Through the above calculation and analysis, we can get $\left| {\sum\limits_{{r} = 0}^{R - 1} {{P_u^m}\left( {t + {r}T} \right)w_L^{  l{r}T}} } \right|= \sqrt R$. 

Additionally, $A$ is a circular Florentine array, it is known that $A^m$ is a permutation on $\mathbb{Z}_T$. And we have  
\[{\left| {{P_u^m}\left( {t+ {r}T} \right)} \right|}=\left| {w_R^{\left( {m + 1} \right)\frac{{{r}\left( {1 + {r}} \right)}}{2}}w_T^{ut}} \right|=1.\]
According to Lemma 5 and 6, any sequence in ${S^m}$ obtained by Theorem 2 is unimodular and perfect. This completes the proof of Part 1). 

We now prove Part 2). Let ${\textbf{s}_u^m}$ and ${\textbf{s}_v^m}$ be any two sequences in $S^m$, where $0 \le u\ne v < T$ and $0 \le m <M$. Based on the IFZT and Lemma 7, for $\tau_1=0$, we consider the following three cases to evaluate (15). 

Case 1: When ${\tau_2}=0$, (15) becomes
\begin{equation}
\begin{aligned}
&\sum\limits_{{r} = 0}^{R - 1} {\sum\limits_{t = 0}^{T - 1} {w_L^{  0\left( {A^m\left( t\right) + {r}T} \right)}{P_u^m}\left( {t+ {r}T} \right)P_v^{m*}\left( {t + {r}T} \right)} }\\
=&\sum\limits_{{r} = 0}^{R - 1}{\sum\limits_{t = 0}^{T - 1} {P_u^m}\left( {t+ {r}T} \right)P_v^{m*}\left( {t + {r}T} \right)} \\=&R\sum\limits_{t = 0}^{T - 1} {w_T^{\left( {u - v} \right)t}}.
\end{aligned}
\end{equation}
Then (32) is equal to zero followed by the fact that $\left( {{u - v}} \right)t$ is a permutation on $\mathbb{Z}_T$ for any $u \ne v$. 

Case 2: When $0 < {\tau_2}< R$, we have
\begin{equation}
\begin{array}{l}
\begin{aligned}
&\sum\limits_{{r} = 0}^{R - 1} {\sum\limits_{t = 0}^{T - 1} {w_L^{  {\tau_2}\left( {A^m\left( t\right) + {r}T} \right)}{P_u^m}\left( {t+ {r}T} \right)P_v^{m*}\left( {t + {r}T} \right)} }\\
=&\sum\limits_{{t} = 0}^{T - 1} {w_L^{  {\tau_2}{A^m}\left( {{t}} \right)}w_T^{\left( {u - v} \right){t}}} \sum\limits_{{r} = 0}^{R - 1} {w_R^{  {\tau_2}{r}} }.
\end{aligned}
\end{array}
\end{equation}
Due to  $\sum\limits_{{r} = 0}^{R - 1} {w_R^{  {\tau_2}{r}}}=0$ for ${\tau_2} \ne 0$, (33) is equal to zero for $0 < {\tau_2}< R$.

Case 3: When ${\tau_2} =R$, (15) becomes
\begin{equation*}
\begin{aligned}
&\sum\limits_{{r} = 0}^{R - 1} {\sum\limits_{t = 0}^{T - 1} {w_T^{  \left( {A^m\left( t\right) + {r}T} \right)}{P_u^m}\left( {t+ {r}T} \right)P_v^{m*}\left( {t + {r}T} \right)} }\\
=&\sum\limits_{{r} = 0}^{R - 1} {{w^{  {r}}}\sum\limits_{{t} = 0}^{T - 1} {w_T^{  A^m\left( {{t}} \right) + \left( {u - v} \right){t}}} }. \\
\end{aligned}
\end{equation*}

Within Construction \uppercase\expandafter{\romannumeral1}, it is established that $A^m\left( {{t}} \right) \ne at$, where $a$ is a constant. This implies that $ A^m\left( {{t}} \right) + \left( {u - v} \right)t$ is neither 0 nor a multiple of $t$. Therefore,  
 $0< \left|\sum\limits_{{r} = 0}^{R - 1} {{w^{  {r}}}\sum\limits_{{t} = 0}^{T - 1} {w_T^{  A^m\left( {{t}} \right) + \left( {u - v} \right){t}}} }\right|<RT$ and $0 < \left| {{\theta _{{\bf{s}}_u^m,{\bf{s}}_v^m}}\left( T  \right) } \right| < RT^2$.

From Cases 1-3, we conclude that each $S^m$ is an optimal $\left( {{RT^2},T,RT} \right) $-ZCZ set respect to the Tang-Fan -Matsufuji bound. Furthermore, all sequences in ${S^m}$ exhibit the property of cyclic distinctness.

Finally, we prove Part 3). Let ${\textbf{s}_u^{m_1}}$ and ${\textbf{s}_v^{m_2}}$ be two sequences in $S^{m_1}$ and $S^{m_2}$, respectively, where $0 \le u\ne v < T$ and $0 \le {m_1} \ne {m_2} <M$. Let $n = t + lT, \tau  = {\tau _1} + {\tau _2}T, 0 \le t,{\tau _1} < T, 0 \le l, {\tau _2} < L$, we have
\begin{equation*}
\label{e:barwq}
\begin{split}
&\left| {\sum\limits_{{r} = 0}^{R - 1} {P_u^{{m_1}}\left( {t + {\tau _1}+{r}T} \right){P_v^{{m_2}*}\left( {t +{r}T} \right)w_R^{ {r}{\tau _2}}} }} \right|\\
=&\left| {w_T^{u({t+ {\tau _1})-vt}}\sum\limits_{{r} = 0}^{R - 1} {w_R^{\left( {{m_1}-{m_2}} \right)\frac{{{r}\left( {1 + r} \right)}}{2}{+r}{\tau _2}}} } \right|\\
= &\left| {\sum\limits_{{r} = 0}^{R - 1} {w_R^{ {r} {\tau _2}}w_R^{\left( {{m_1} - {m_2}} \right)\frac{{{r}\left( {1 + {r}} \right)}}{2}}} } \right|.
\end{split}
\end{equation*}

We observe that the above  equation shares a similar structure with (31). This allows us to exploit an analogous approach and conclude that
\begin{equation*}
\left| {\sum\limits_{{r} = 0}^{R - 1} {w_R^{ {r}{\tau _2}}w_R^{\left( {{m_1} - {m_2}} \right){\frac{{{r}\left( {1 + {r}} \right)}}{2}} }} } \right|=\sqrt R.
\end{equation*}

According to Lemma 8, we have $\left| {{\theta _{{\bf{s}}_u^{{m_1}},{\bf{s}}_v^{{m_2}}}}\left( \tau  \right)} \right| = \sqrt R T$ for all $0 \le {m_1} \ne {m_2} <M, 0 \le u\ne v < T$ and $0 \le \tau  < R{T^2}$. 
\hfill
$\hfill\blacksquare$ 
\section*{Proof of the Theorem 3}
\textit{Proof:} For $R$ being even, $0 \le t<T$ and $0 \le l<L$, we have
\begin{equation}
\begin{aligned}
&\left|{\sum\limits_{{r} = 0}^{R - 1} {{P_u}\left( {t + {r}T} \right)w_L^{  l{r}T}} } \right|^2\\
 =& \left( {\sum\limits_{{r} = 0}^{R - 1} {w_{2R}^{{r}^2}w_T^{ut}w_L^{  l{r}T}} } \right)\left( {\sum\limits_{{r'} = 0}^{R - 1} {w_{2R}^{ - {r'}^2}w_T^{  -ut}w_L^{-l{r'}T}} } \right)\\
 =& \left( {\sum\limits_{{r} = 0}^{R - 1} {w_{2R}^{{r}^2}w_L^{l{r}T}} } \right)\left( {\sum\limits_{{r'} = 0}^{R - 1} {w_{2R}^{-{r'}^2}w_L^{-l{r'}T}} } \right)
\end{aligned}
\end{equation}

By analyzing the values of  $\sum\limits_{{r} = 0}^{R - 1} {w_{2R}^{{r}^2}w_L^{  l{r}T}}$ for different values of $r$, (34) can be reduced to

\begin{equation}
\left|{\sum\limits_{{r} = 0}^{R - 1} {{P_u}\left( {t + {r}T} \right)w_L^{  l{r}T}} } \right|^2
={\sum\limits_{{r} =0}^{R - 1} {w_{2R}^{{r}^2 }} } {\sum\limits_{{r'} = 0}^{R - 1} {w_{2R}^{-{r'}^2 }} } .
\end{equation}

Let $a = \frac{R}{2}$, through further analysis, (35) can be reformulated as
\begin{equation*}
\begin{aligned}
&\left( {\sum\limits_{{r} = 1}^{a - 1} {w_{2R}^{{r}^2}} +w_{2R}^{0 }+w_{2R}^{{a^2}}} \right)\left( {\sum\limits_{{r'} = 1}^{a - 1} {w_{2R}^{-{r'}^2}} +w_{2R}^{0 }+w_{2R}^{-{a^2}}} \right)\\
&=  {\sum\limits_{{r} = 1}^{a - 1} {w_{2R}^{{r}^2}} }  {\sum\limits_{{r'} = 1}^{a - 1} {w_{2R}^{-{r'}^2}} }  +\left( {w_{2R}^{0 }+w_{2R}^{{a^2}}} \right) {\sum\limits_{{r'} = 1}^{a - 1} {w_{2R}^{-{r'}^2}} } \\
&+\left( {w_{2R}^{0 }+w_{2R}^{-{a^2}}} \right){\sum\limits_{{r} = 1}^{a - 1} {w_{2R}^{{r}^2}} }+ \left( {w_{2R}^{0 }+w_{2R}^{{a^2}}} \right)\left( {w_{2R}^{0 }+w_{2R}^{-{a^2}}} \right)\\
 &= 4\left( {a - 1} \right) + 2 - 2\left( {a - 1} \right)=R.
\end{aligned}
\end{equation*}

Then, we get $\left| {\sum\limits_{{r} = 0}^{R - 1} {{P_u}\left( {t + {r}T} \right)w_L^{  l{r}T}} } \right|= \sqrt R$. 

Additionally, $A$ is a permutation on $\mathbb{Z}_T$. And we have  
\[\left| {{P_u}\left( {t + {r}T} \right)} \right|  = \left| {w_{2R}^{{r}^2}w_T^{ut}} \right| =1.\]
We assert that any sequence in ${S}$ is unimodular and perfect.  This completes the proof of Part 1). 

As the demonstration here closely parallels the approach used in Theorem 2 Part 2), a detailed proof is skipped to avoid redundancy. 
}

\end{document}